\documentclass[fleqn,usenatbib]{mnras}
\usepackage{fix-cm}
\usepackage{lscape}
\usepackage{newtxtext,newtxmath}
\usepackage{threeparttable} 

\usepackage[T1]{fontenc}
\usepackage{makecell}
\DeclareRobustCommand{\VAN}[3]{#2}
\let\VANthebibliography\thebibliography
\def\thebibliography{\DeclareRobustCommand{\VAN}[3]{##3}\VANthebibliography}

\usepackage{graphicx}	
\usepackage{amsmath}	
\usepackage{float}

\title[Sco X-1: XSPECT]{Spectral nature of Sco X-1 observed using the X-ray SPECtroscopy and Timing (XSPECT) payload on-board XPoSat}

\author[Shyam Prakash et al.]{V. P. Shyam Prakash$^{1,2}$\thanks{E-mail: shyamvp151@gmail.com},
Vivek K. Agrawal$^{1}$,
Rwitika Chatterjee$^{1}$,
Radhakrishna Vatedka$^{1}$,
\newauthor
Koushal Vadodariya$^{1}$ 
and A. M. Vinodkumar$^{2}$
\\
$^{1}$Space Astronomy Group, ISITE Campus, U. R. Rao Satellite Center, ISRO, Bengaluru 560037, India\\
$^{2}$Department of Physics, University of Calicut, Kerala, 673635, India
}

\date{Accepted XXX. Received YYY; in original form ZZZ}

\pubyear{\the\year{2025}}

\begin{document}
\label{firstpage}
\pagerange{\pageref{firstpage}--\pageref{lastpage}}
\maketitle

\begin{abstract}
Scorpius X-1 is the brightest and first discovered X-ray source in the sky. Studying this source in the low-energy band has been challenging in the past due to its high brightness. However, with the X-ray SPECtroscopy and Timing (\textit{XSPECT}) payload on-board India’s first X-ray Polarimetry Satellite (\textit{XPoSat}), we have the capability to study the source despite its very high brightness, thanks to the fast ($\sim$1 ms) readout of the instrument. We investigate the evolution of the spectral and timing properties of Sco X-1 across the horizontal, normal, and flaring branch, as observed with XSPECT. We examine changes in the spectral parameters as a function of position on the color-color diagram (CCD). Spectral studies indicate that the soft X-ray emission can be modeled using a multicolor disk component, with the inner disk temperature ranging from $\sim$0.6 to 0.8 keV. The hard component is described by a Comptonized continuum using either the \textit{nthComp} or \textit{Comptb} model with electron temperatures from $\sim$2.4 to 4.7 keV and optical depth between $\sim$5 and $\sim$14. Additionally, we observe the presence of an iron $K_\alpha$ line at $\sim$6.6 keV and an iron $K_\beta$ line at $\sim$7.6 keV. Both spectral models suggest a steep rise in Comptonization flux as well as disk flux in the flaring branch. An increase in neutron star blackbody temperature and inner disk temperature are also observed during flaring. The Z-track is driven by changes in the optical depth of the corona, the Comptonization flux and the disk flux and the inner disk temperature. No quasi-periodic oscillations are detected in any branch, suggesting their association with the high-energy spectrum.

\end{abstract}

\begin{keywords}
accretion, accretion disks – stars: individual: Scorpius X–1 – stars: neutron – X-rays: binaries – X-rays: general
\end{keywords}



\section{Introduction}\label{intro}
In a neutron star low-mass X-ray binary system (NS-LMXB), matter is being accreted from a low-mass ($< 1M_{\odot}$) companion star to a weakly magnetized neutron star (NS), via Roche-lobe overflow through the inner Lagrange point (\citealt{2006csxs.book..623T}).
Based on its evolutionary pattern they trace on the Hardness-intensity diagram (HID) or Color–color diagram (CCD), NS-LMXBs are classified into `Z' and `atoll' sources and their spectral and temporal properties are strongly correlated with the position on the track (\citealt{1989A&A...225...79H}).
In Z-sources, the Z-track is split into three branches (horizontal (HB), normal (NB) and flaring branch (FB)) and atoll sources are found in high soft state (otherwise called the `banana state') or low hard state (`island state').
There are a few sources which exhibit both Z and atoll behavior.
For example, \citealt{2009ApJ...696.1257L} has reported an evolution of the transient source, XTE J1701-462, from a Cyg-like Z-source to a Sco-like Z-source and to an atoll-source during its 2006-2007 outburst.
IGR J17480–2446 (\citealt{2010ATel.2952....1A}), XTE J1701-462 (\citealt{2010ApJ...719..201H}), XTE J1806-246 and Cir X-1 have also shown the transition in behavior between Z and atoll nature.
Z-sources are generally high luminosity systems (L $\sim$ $L_{Edd}$), whereas atoll-source exhibit a much wider range of luminosity from $\sim$ 0.01-0.5$L_{Edd}$.
The timescale of evolution of the sources through the Z-track is few hours to days and this evolution could be driven by the change in mass accretion rate ($\dot{M}$) across the different parts of the Z-track (\citealt{1990A&A...235..131H}).
The process driving the movement of source along the `Z' and `atoll' tracks are still debated.

The continuum in the X-ray spectra of NS-LMXBs is described as a sum of a soft thermal component which dominates in lower energies ($<$ 1 keV) and a Comptonized hard component.
There are mainly two widely accepted explanations for this behavior.
The first one considers the soft component to be the emission from a multicolor standard accretion disk and the hard component resulting from inverse Compton scattering of the soft seed photons by hot plasma in the boundary layer (BL; \citealt{2001AIPC..599..870P}) or corona close to the NS surface.
This approach has been widely used in the past for describing the X-ray spectra in various sources \citep{1984PASJ...36..741M, 2002A&A...386..535D, 2009MNRAS.398.1352A, 2023MNRAS.518..194A}.
On the other hand, in the second approach, the soft thermal component is attributed to a single temperature blackbody emission from the boundary layer or the NS surface, and the hard component is attributed to the hot inner accretion flow \citep{1988ApJ...324..363W, 2016ApJ...827..134S}.
Apart from the above two scenarios, a combination of two blackbody models has also been used to describe the continuum in X-ray spectra of NS-LMXB systems, comprising a multicolor disk emission and a blackbody emission from the neutron star surface or boundary layer \citep{2012ApJ...756...34L, 2023MNRAS.523.2788K}.
In addition to the soft and hard components mentioned above, the presence of reflection features have been observed in various NS-LMXB sources \citep{2008ApJ...674..415C, 2017ApJ...836..140L}.

Scorpius X–1 (henceforth, Sco X-1), the brightest X-ray source in the sky, is a persistent, weakly magnetized NS-LMXB. 
It was the first extrasolar X-ray source discovered in 1962 (\citealt{1962PhRvL...9..439G}), with observed luminosity($L$) close to the Eddington limit($L_{Edd}$).
A peak luminosity of around $2 \times 10^{38} \, \text{erg} \, \text{s}^{-1} $ is observed in the source, which corresponds to a luminosity close to the Eddington limit for a 1.4 $M_{\odot}$ neutron star (\citealt{2014ApJ...789...98T}).
The companion in the binary system is found to be an M-type star with a mass $\sim$ 0.4 $M_{\odot}$ (\citealt{2002ApJ...568..273S}).
The orbital period of $\sim$ 19 hrs for the binary system was first estimated by \citealt{1975ApJ...195L..33G} using observations in the Optical band.
Sco X-1 is located at a distance of 2.8 $\pm$ 0.3 kpc, as measured from the trigonometric parallax of the source using Very Long Baseline Array(\citealt{1999ApJ...512L.121B}). Later, the Giaia Data Release 2 derived the distance to the source to be 2.13 $^{+0.26}_{-0.21}$ kpc (\citealt{2021MNRAS.502.5455A}).
The source is found to have an inclination of $\sim$ 25$^{\circ}$ - 34$^{\circ}$ (\citealt{2022aems.conf..134C}). 
It is also the first X-ray binary from which radio emissions were detected (\citealt{1968Natur.218..855A}).
The Jet observed in Sco X-1 using the Very long baseline interferometry observations revealed a position angle of $\sim $54$^{\circ}$ and inclination 44$^{\circ}$ $\pm$ 6$^{\circ}$ \citep{2001ApJ...553L..27F, 2001ApJ...558..283F}.
Energy-dependent polarization was first observed in Sco X-1 by PolarLight satellite (\citealt{2022ApJ...924L..13L}).
A highly significant detection of polarization from the source was observed with \textit{IXPE} at at polarization degree (PD) of 1.0\% $\pm$ 0.2\% and a polarization angle (PA) of 8$^{\circ}$ $\pm$ 6$^{\circ}$ (\citealt{2024ApJ...960L..11L, 2021A&A...654A.102M}).
Sco X-1 is classified as a Z source based on the track it creates in the Color-Color diagram (CCD).
It exhibits a Z-track with an extended FB. 
This is because of the frequent flaring events occurring in the system with a higher X-ray flux.

The energy spectra of Sco X-1 are modeled using a combination of a thermal component either from the inner parts of the accretion disk or from the NS surface, an iron emission line and the Comptonized component \citep{2003A&A...405..237B,2012A&A...546A..35C, 2014ApJ...789...98T}. 
The presence of $K_{\alpha}$ and $K_{\beta}$ emission lines of the
He-like Fe xxv ions were detected in the NuSTAR spectrum (\citealt{2021A&A...654A.102M}). 
Also, a hard X-ray tail was detected in the source using HXMT observation (\citealt{2023ApJ...950...69D}).
QPOs are observed in normal branch (NBO) at $\sim$ 6 Hz, horizontal branch at $\sim$ 45 Hz of the Z-track along with kHz QPOs (\citealt{1996ApJ...469L...1V}).
kHz QPOs were discovered in Sco X-1 using RXTE observation by \citealt{1996ApJ...469L...1V}.
Later, a study of QPOs in Sco X-1 by \citealt{2021ApJ...913..119J} notes that no QPOs were observed in either the HB or NB below 6 keV associating the origin of QPOs with the hard Comptonization component.
All of these studies were conducted in the high energy band (above 3 keV) due to extremely high brightness of the source in the low-energy band, which can saturate the detectors in the telescope. 

India's first X-ray polarimetric mission, XPoSat was launched on January 1, 2024. 
The X-ray SPECtroscopy and Timing (\textit{XSPECT}: \citealt{2025arXiv250520061V}) payload on-board XPoSat provides spectroscopic and timing information in the 0.8-15.0 keV energy range along with Polarimeter Instrument in X-rays (\textit{POLIX}; \citealt{2016cosp...41E1533P}) payload.
\textit{XSPECT} uses an array of Swept Charge Devices (SDDs) to provide long-term monitoring of changes in the spectral state and simultaneous long term temporal monitoring of soft X-ray emission.
The instrument has an effective area of $>$30 cm$^{2}$ with an energy resolution of $\sim$180 eV at 6 keV.
A total of 16 SDDs are arranged in four quads with two out of the four quads provide a Field of View (FoV) of $3^{\circ} \times 3^{\circ}$ and the rest with $2^{\circ} \times 2^{\circ}$ FoV.
One detector is covered with Tantalum sheet to prevent X-rays and this detector act as charge particle monitor.
With the fast readout of 1 ms of the instrument, \textit{XSPECT} can handle high count rates and thus can provide pile-up free observational data even for bright sources in the soft X-ray energy range (0.8-15.0 keV).

The paper is organized as follows; The details of \textit{XSPECT} observations of Sco X-1 and the data reduction procedure are given in Section \ref{obs}. Section \ref{data_analyisis} deals with the spectral and timing data analysis. The results obtained from spectral and timing analysis are presented in \ref{res}. Finally, we discuss the results of the study in Section \ref{sum}. 

\section{Observations and Data reduction}\label{obs}

\begin{table}
    \centering
    \begin{threeparttable}
    \begin{tabular}{lccc}
    \hline
    \hline
     Observation ID    &  \makecell{Observation date\\ (DD-MM-YYYY)} &  MJD start &  Exposure (s) \\
     \hline
         & 04-08-2024 & 60526 & 13153.0  \\
         & 05-08-2024 & 60527 & 13644.0  \\
         & 06-08-2024 & 60528 & 12408.0  \\
         & 07-08-2024 & 60529 & 12620.0  \\
         & 16-08-2024 & 60538 & 13653.0  \\
         & 17-08-2024 & 60539 & 13165.0  \\
    X01\_XPC\_G01\_0006  & 18-08-2024 & 60540 & 12809.0  \\
         & 19-08-2024 & 60541 & 12814.0  \\
         & 20-08-2024 & 60542 & 13594.0  \\
         & 21-08-2024 & 60543 & 12238.0  \\
         & 22-08-2024 & 60544 & 12810.0  \\
         & 23-08-2024 & 60545 & 720.0\tnote{1} \\
         & 24-08-2024 & 60546 & 12194.0  \\
         & 25-08-2024 & 60547 & 9628.0  \\
         & 26-08-2024 & 60548 & 10971.0 \\
    \hline
    \hline
    \end{tabular}
    \begin{tablenotes}
    \item[1] Low exposure due to partial data received.
    \end{tablenotes}
    \end{threeparttable}
    \caption{Log of \textit{XSPECT} observations of Sco X-1.}
    \label{log}
\end{table}

\textit{XSPECT} observed Sco X-1 several times (15 days) in August 2024 (Obs ID: \texttt{X01\_XPC\_G01\_0006}) with an effective total exposure time of approximately 176 ks. 
The detailed observation log is presented in Table \ref{log}.
We used these observations to study the spectral behavior of the source and its evolution through the Z-track.
The day-wise Level-1 (L1) files were processed using the \texttt{xspl2screen} task as part of the \textit{XSPECT pipeline} (Will be made available through the Indian Space Science Data Center (ISSDC)\footnote{\href{https://xpps.issdc.gov.in/web/}{https://xpps.issdc.gov.in/web/}}).
The script generates the Level-2 (L2) event files, after incorporating the Good Time Intervals (GTIs).
The individual day-wise observational files are merged using the \texttt{xspevtmerge} task.
Science products such as the spectra and light curve are generated using the \texttt{xspl2specgen} and \texttt{xspl2lcgen} tasks respectively.
The pipeline also provide option to generate detector wise spectral and light curves.
Spectra and light curve are created separately for both the $2^{\circ} \times 2^{\circ}$ and $3^{\circ} \times 3^{\circ}$ FoV detectors.
The individual detector light curves are then added together to create the source light curve.
However, the spectra from the $2^{\circ} \times 2^{\circ}$ and $3^{\circ} \times 3^{\circ}$ FOV detectors are considered independently for spectral studies. 

\begin{figure*}
    \centering
    \includegraphics[width=0.55\linewidth]{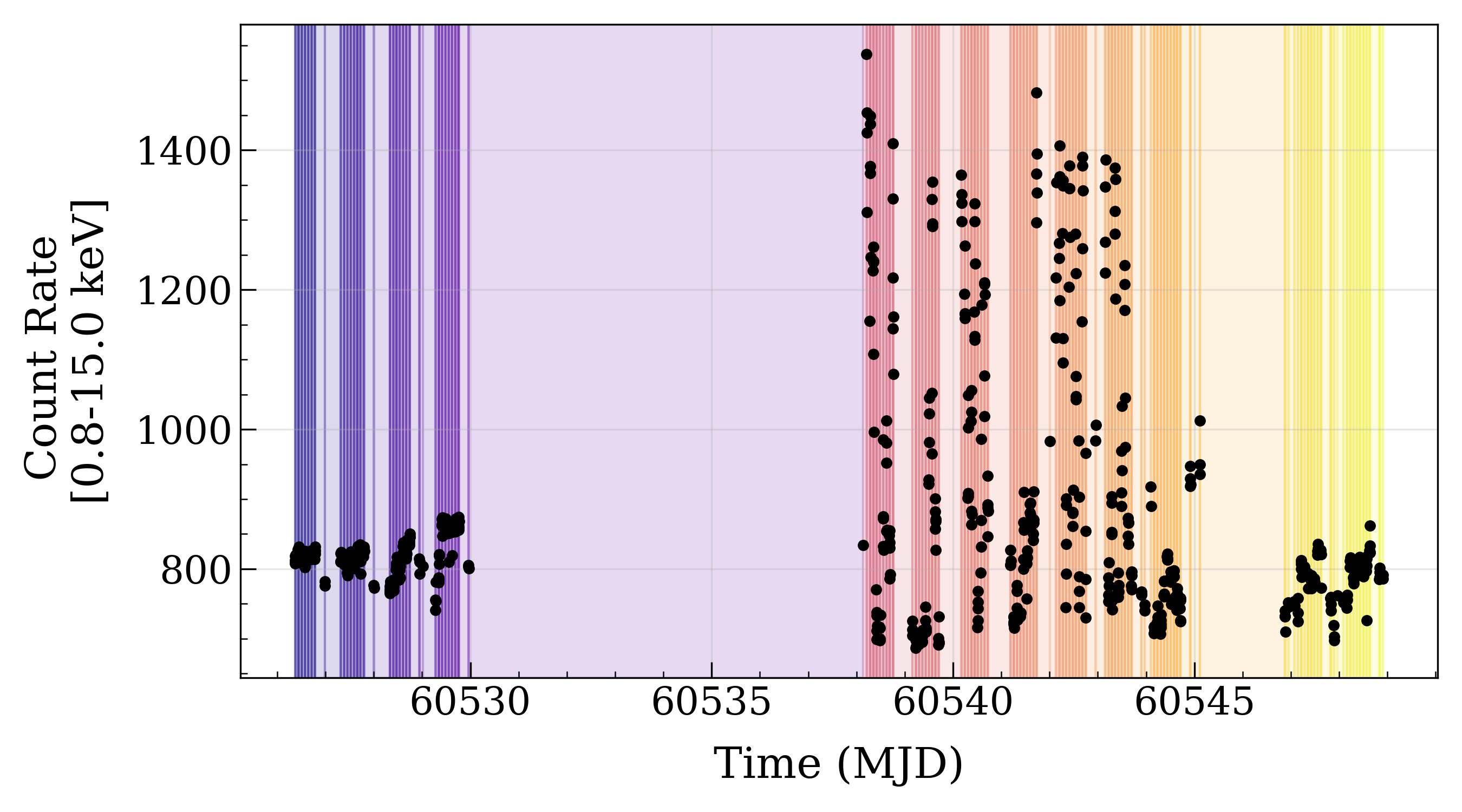}
    \includegraphics[width=0.43\linewidth]{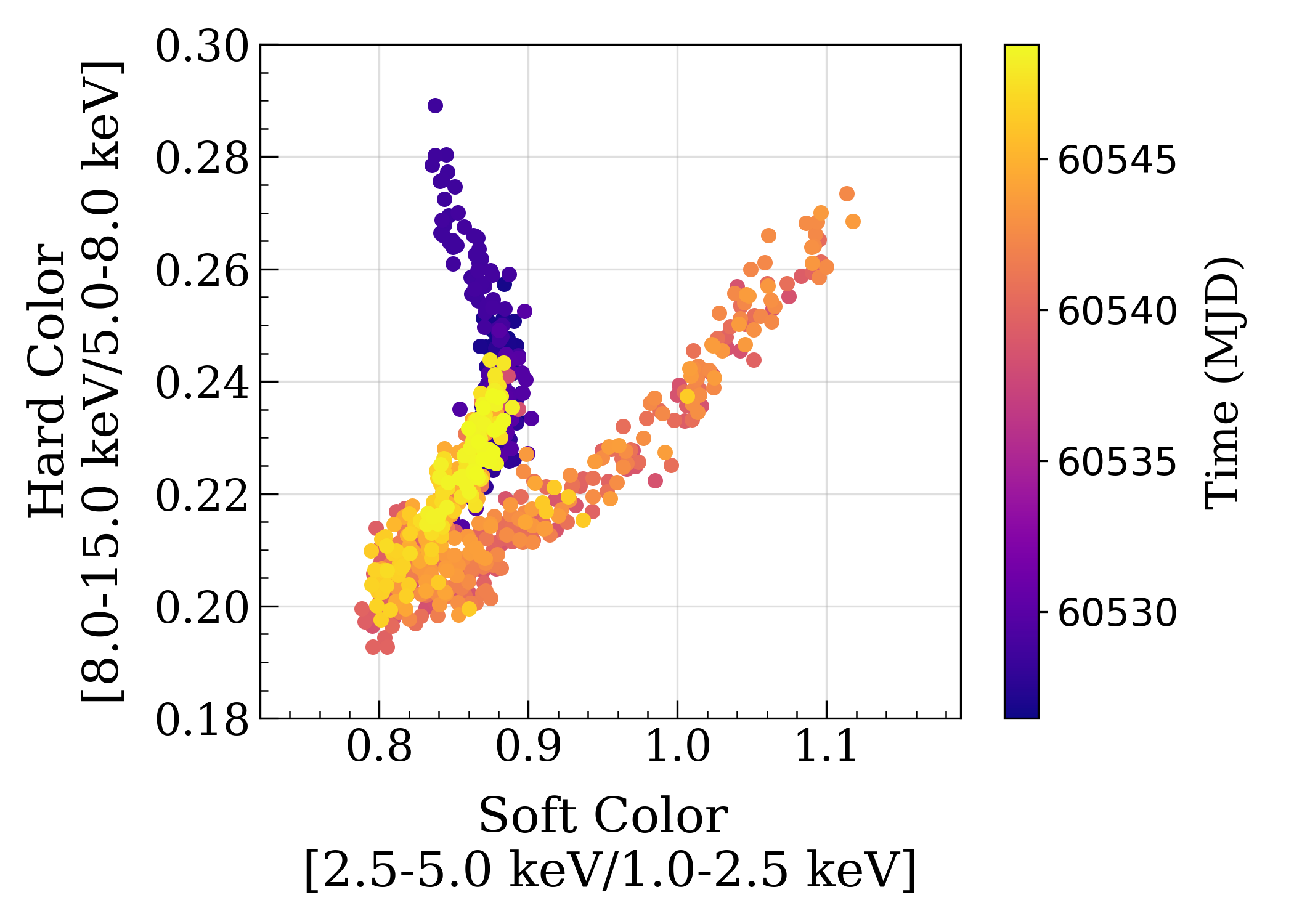}
    \caption{(Left) \textit{XSPECT} light curve of Sco X-1 plotted for a bin size of 200 sec in the 0.8-15 keV energy band. Flaring activities are observed in the source during the observation in 60538-60545 (MJD). (Right) HID plotted using \textit{XSPECT} observations of Sco X-1. Each data point represents a 200 s bins. Color code is given to each point to demonstrate the evolution of source through the Z track during the observation.}
    \label{hid}
\end{figure*}

\section{Data analysis}\label{data_analyisis}

\subsection{Light curve and Hardness intensity diagram}

The \textit{XSPECT} light curve of Sco X-1 in the 0.8-15.0 keV during the observation is shown in the left panel of Figure \ref{hid} with X-axis displaying the days in MJD. 
The light curve is plotted in 200 s bins by summing both the $2^{\circ} \times 2^{\circ}$ and $3^{\circ} \times 3^{\circ}$ FoV detector light curves.
The source is seen to exhibit sudden flux rise (flaring) in its light curve close to the middle of the observations (MJD: 60322-60329).
The maximum intensity ($\sim$ 1600 counts per second) during these events goes to twice the nominal value at the beginning of the observation.
The right panel of Figure \ref{hid} shows the CCD plotted for the observations with each point representing a 200 sec bin.
A time-coded color is given to each data point to understand the evolution of source in its CCD during the observations.
The soft color is defined as the ratio of counts in 2.5-5.0 keV and 1.0-2.5 keV energy band, whereas the hard color is defined as the ratio of 8.0-15.0 to 5.0-8.0 keV counts.
The CCD clearly points out the three well-defined segments of the Z-tracks, corresponding to the horizontal, normal and flaring branches of the Z-track and two apex (soft apex (SA) and hard apex (HA)).
The color of vertical background lines displayed in the light curve are correlated with the color of data point in the color-color diagram (CCD) for comparison.

\subsection{Spectral analysis}

\begin{figure}
    \centering
    \includegraphics[width=0.9\linewidth]{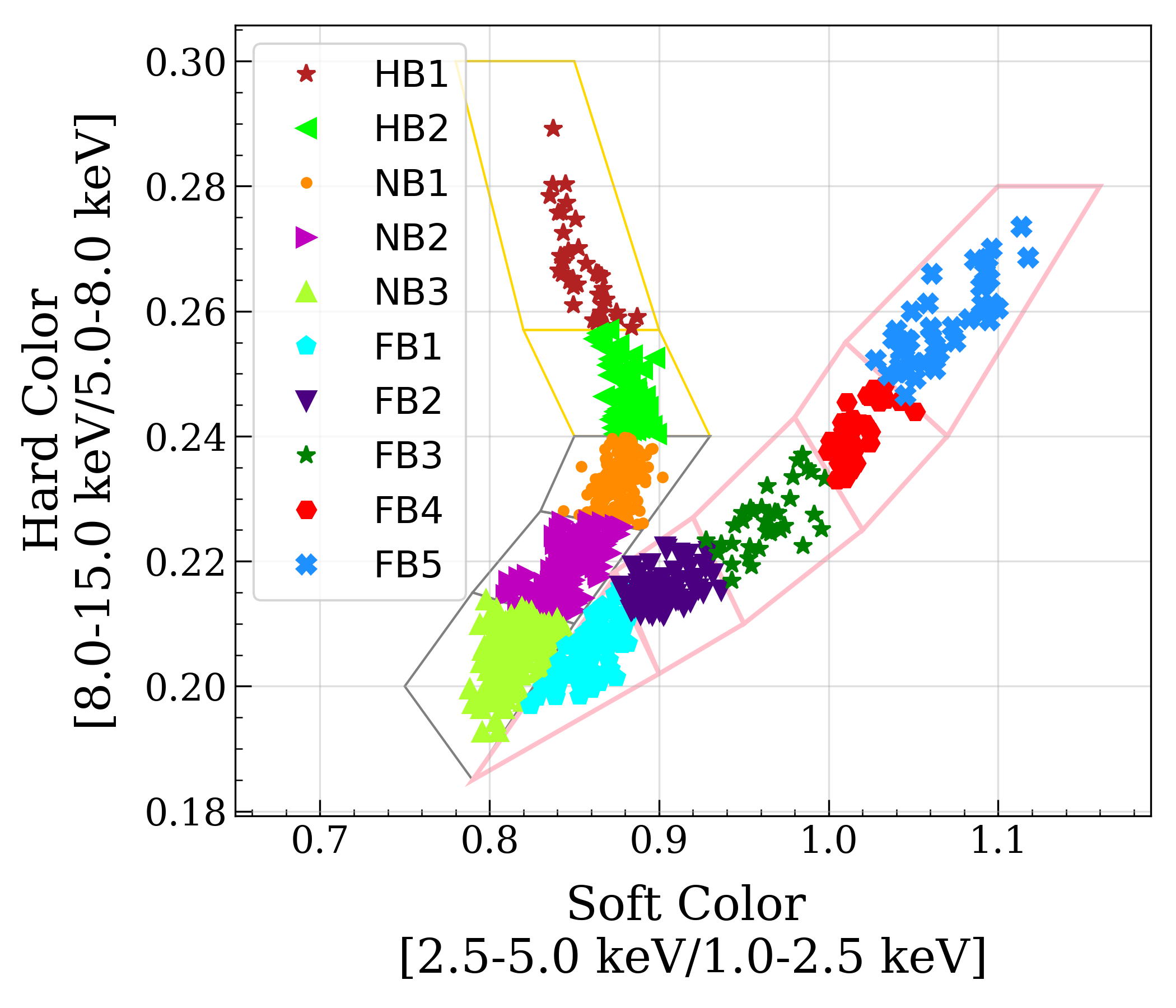}
    \caption{HID plotted using \textit{XSPECT} observations of Sco X-1 for a 200 s bins. The Z-track is divided into seven segments for spectral and timing studies.}
    \label{hid2}
\end{figure}

Spectral fitting is performed in the energy range of 0.8-15.0 keV using \texttt{XSPEC}.V12 (\citealt{1996ASPC..101...17A}).
The Z-track is divided into ten segments (two segments in HB, three segments in NB, five segments in FB), as shown in Figure \ref{hid2}.
The segments are considered in such a way that all the segments have comparable exposure.
Each point in the plot represents a 200 sec bin size. The GTIs corresponding to the data points in each of the segments in the CCD are identified. 
These GTIs are then used to extract the source spectra for each 200 sec exposure.
The spectra corresponding to these GTIs are then merged to create the source spectra for a particular segment.
The spectra are created separately for $2^{\circ} \times 2^{\circ}$ and $3^{\circ} \times 3^{\circ}$ FOV detectors.
We have binned both the spectra in all segments using the optimal binning scheme by Kaastra \& Bleeker (2016) using \texttt{ftgrouppha} task in \texttt{HEASoft} to have a minimum signal-to-noise ratio (S/N) of 3 per bin. 
The latest background and response files (v.20250319) were employed for the analysis.
A systematic of 1\% is applied to the data, to avoid high residuals in the 1.2-1.8 keV energy range (\citealt{2025arXiv250609918C}). 
The Tubingen-Boulder model (\textit{tbabs} model) is included in the spectral fit to take into account interstellar absorption along the line of sight. 
Initially the value of the equivalent hydrogen column ($N_{H}$) was kept free during the spectral fitting and fixed to average best-fit value of $0.14\times 10^{22} \text{cm}^{-2}$, when not able to constrain. 
The obtained values of $N_{H}$ is consistent with the previously used value for the source (\citealt{2024ApJ...960L..11L})

For bright sources, the inclusion of two edge components, corresponding to
the K-absorption edges of Al and Si are recommended (\citealt{2025arXiv250609918C}).
Hence, we add two edge components at $\sim$ 1.5 and $\sim$ 1.75 keV during spectral fitting to account for the Aluminum and Silicon edges present in the spectra. The response files will be improved in future releases, properly accounting for all these effects.
The FOV $2^{\circ} \times 2^{\circ}$ and $3^{\circ} \times 3^{\circ}$ spectra are treated separately for simultaneous spectral fitting.
The \texttt{gain fit} command is used during spectral fitting to account for the gain correction.
A cross calibration factor is included in the fit and the constant is fixed to 1 for $2^{\circ} \times 2^{\circ}$ and allowed to vary freely for $3^{\circ} \times 3^{\circ}$.

We employed multiple approaches to model the 0.8-15.0 keV \textit{XSPECT} spectra of Sco X-1.
In the first approach (hereafter Model-1), we used a combination of a soft multicolor disk blackbody model (\textit{diskbb}; \citealt{1984PASJ...36..741M}) and a thermal Comptonization model (\textit{nthComp}; \citealt{1996MNRAS.283..193Z}) with blackbody seed photons. This model along with two Gaussian lines at $\sim$ 6.7 keV $\sim$ 7.6 keV yielded a good fit to the spectra across all segments of the Z-track, with reduced $\chi^2$ values close to 1.
In the second approach, a single-temperature blackbody component (\textit{bbodyrad}) replaced the \textit{diskbb} component, while the seed photons for the \textit{nthComp} component were set to originate from the disk. 
This model provided a good statistical fit to the continuum emission across all segments, with a best-fit blackbody temperature ranging between $\sim$ 1.4 keV to 1.8 keV. The disk seed photon temperature ranges between $\sim$ 0.5 keV to 0.8 keV. 
Further (Model-3), we explored a double-Comptonization model comprising two \textit{nthComp} components, with seed photons originating from the disk for the first component and from blackbody emission for the second. 
We tied the electron temperature for both the \textit{nthComp} components.
The spectral fit provided disk photon temperature of $\sim$ 0.8 keV and blackbody seed photon temperature of $\sim$ 1.5 keV.
However, we could not constrain the photon index ($\Gamma$) of the \textit{nthComp} component with blackbody seed photon within the fit for all the segments.
In the fourth approach (Model-4), we use a combination of a multi-color disk blackbody component (\textit{diskbb}) along with a single-temperature blackbody (\textit{bbodyrad}) component to model the source spectra.
However, this model failed to provide an adequate fit across all segments and yielded a $\chi^2/dof > 2$ in all the segments.
Finally, in Model-5, we adopted the Comptonization model \textit{Comptb} (\citealt{2008ApJ...680..602F}) alongside the \textit{diskbb} model. This configuration also yielded good fits with reduced $\chi^2$ values near 1 across all segments.
Among the five models tested, Model-1, Model-2 and Model-5, both involving \textit{diskbb} and \textit{bbodyrad} for soft emission and either \textit{nthComp} or \textit{Comptb} for Comptonized emission provided a good description of the source spectra in all segments along the Z-track.
All these models required additional Gaussian features at $\sim$6.6 and $\sim$7.6 keV, along with the continuum emission to model the $K_\alpha$ and $K_\beta$ transition lines of Fe XXV.
The details of the spectral fit using these models are described in detail.

\begin{figure}
    \centering
    \includegraphics[width=0.85\linewidth]{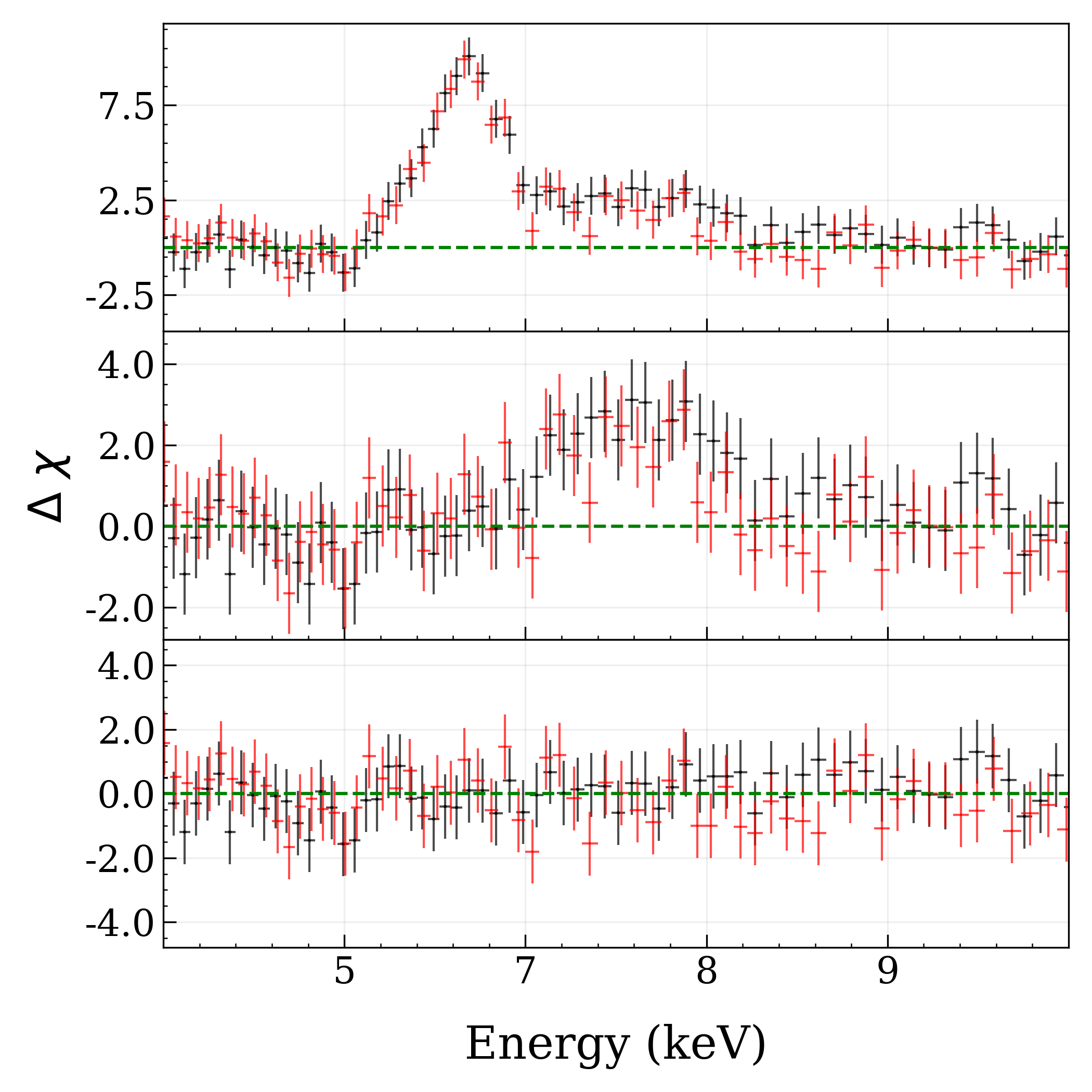}
    \caption{Comparison between the residuals obtained for (top) \textit{tbabs * (diskbb + nthComp)} (middle) \textit{tbabs * (diskbb + nthComp + Gauss)} and (bottom) \textit{tbabs * (diskbb + nthComp + Gauss + Gauss)} showing the presence of Fe $K_{\alpha}$ and $K_{\beta}$ emission lines. The spectra from FoV 2$\times$2 and 3$\times$3 detectors are shown in black and red colors respectively. }
    \label{residual_plot}
\end{figure}

\begin{figure*}
    \centering
    \includegraphics[width=0.33\linewidth]{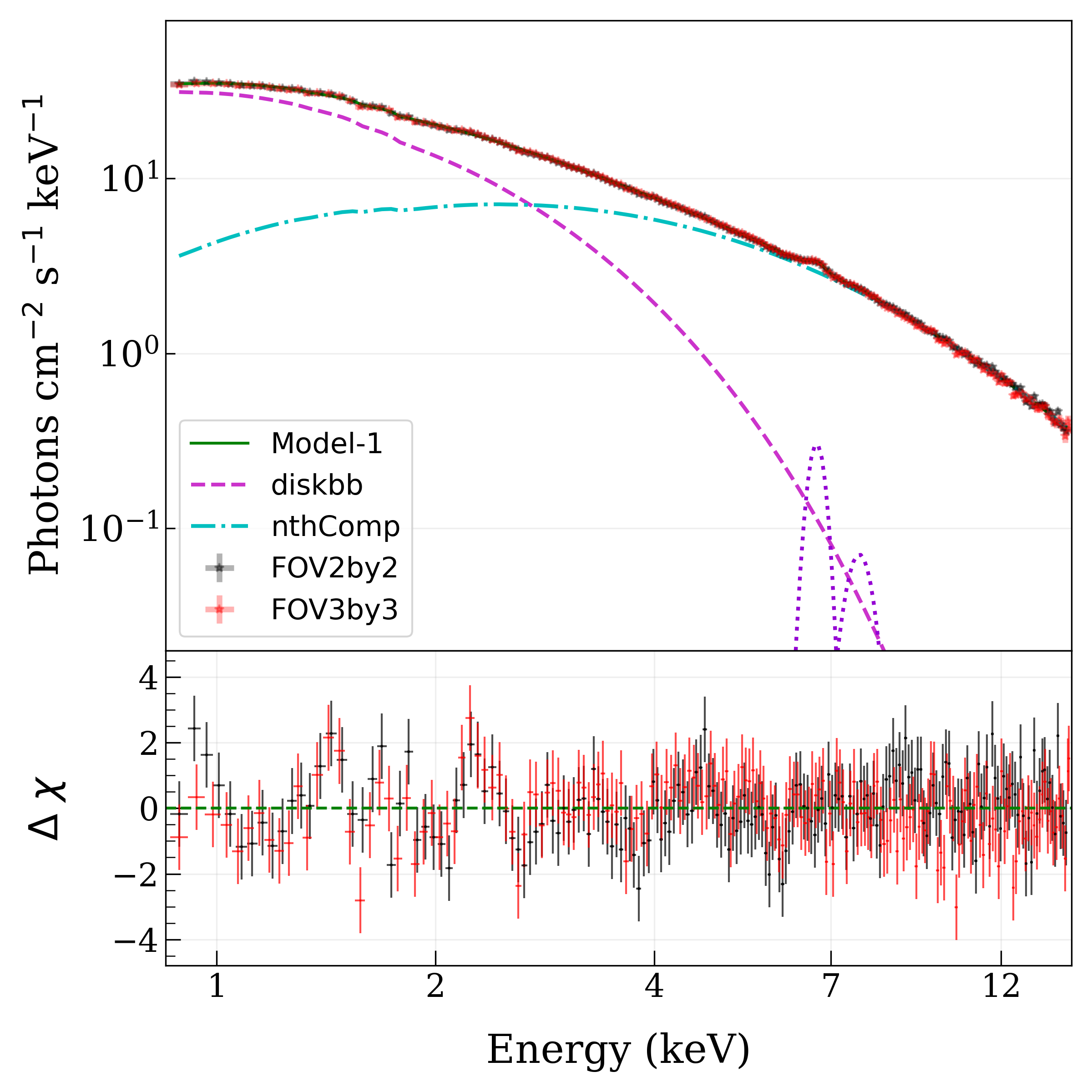}
    \includegraphics[width=0.33\linewidth]{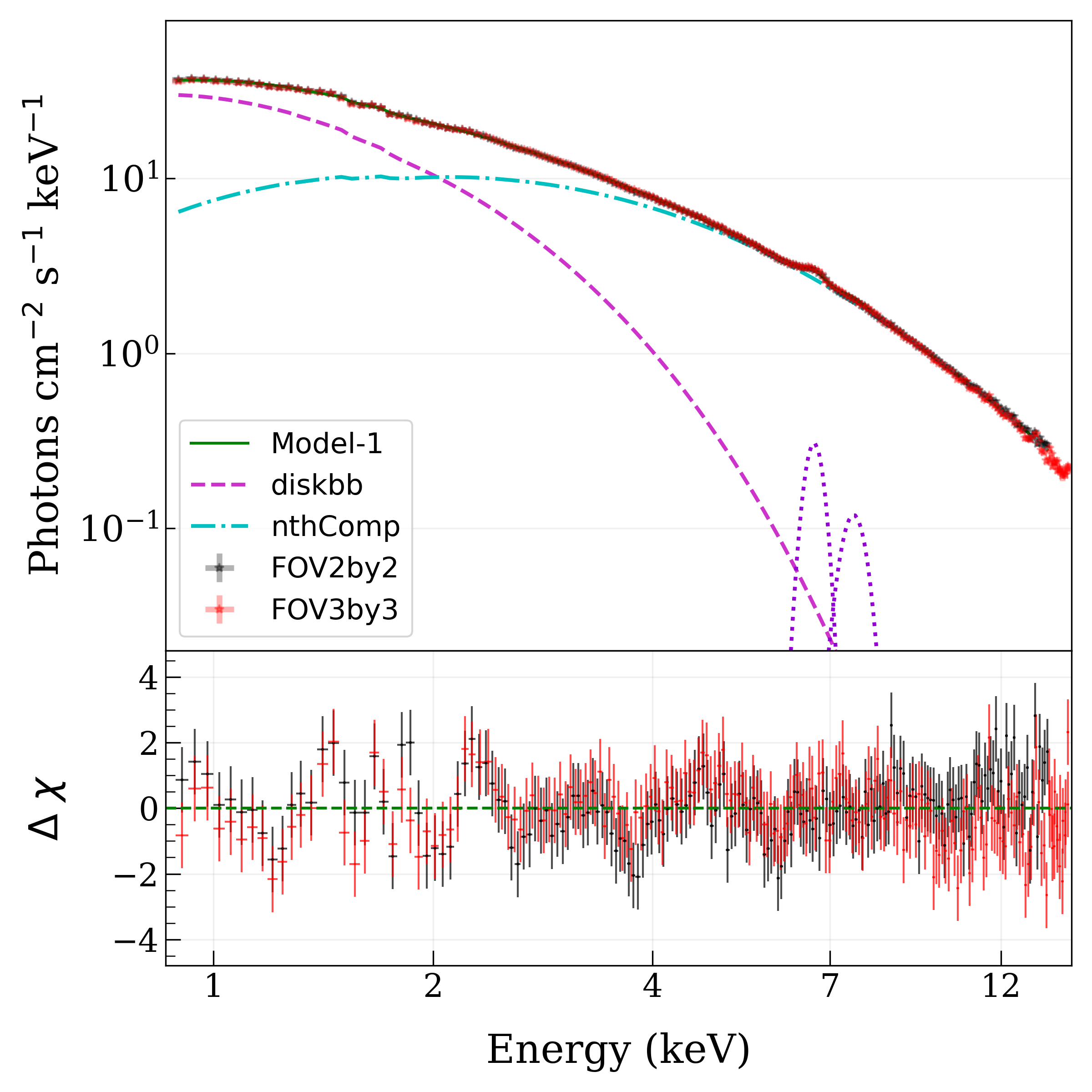}
    \includegraphics[width=0.33\linewidth]{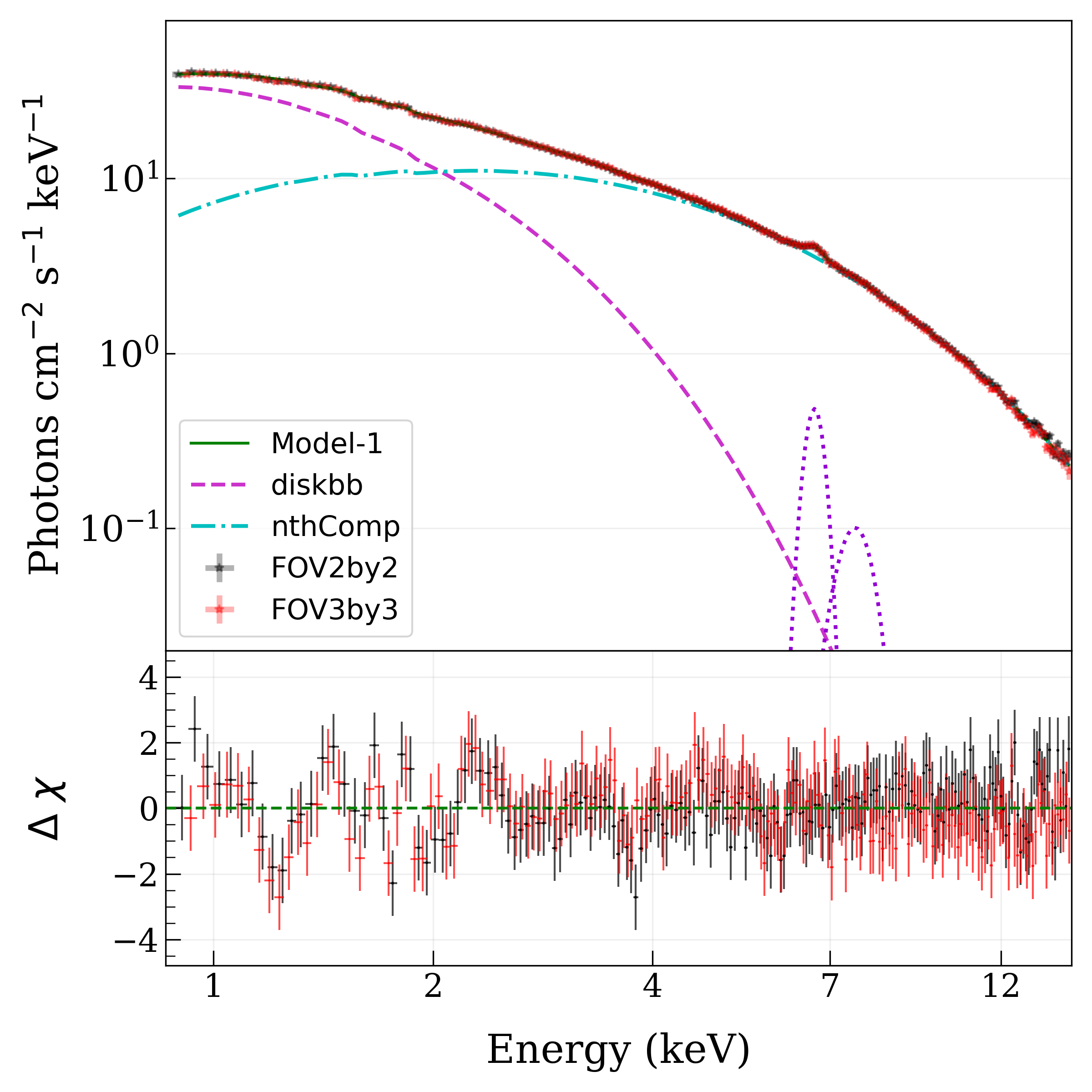}
    \caption{The 0.8-15.0} keV \textit{XSPECT} spectrum of Sco X-1 fitted using Model-1: \textit{tbabs*(diskbb+nthComp+gauss+gauss)} in the HB1 (left), NB1 (middle), and FB1 (right) in \texttt{XSPEC}. The residuals are plotted in the bottom panel. The different components in the model are also plotted along with the model (green solid line).
    \label{spec_fit_model1}
\end{figure*}

\begin{figure*}
    \centering
    \includegraphics[width=0.33\linewidth]{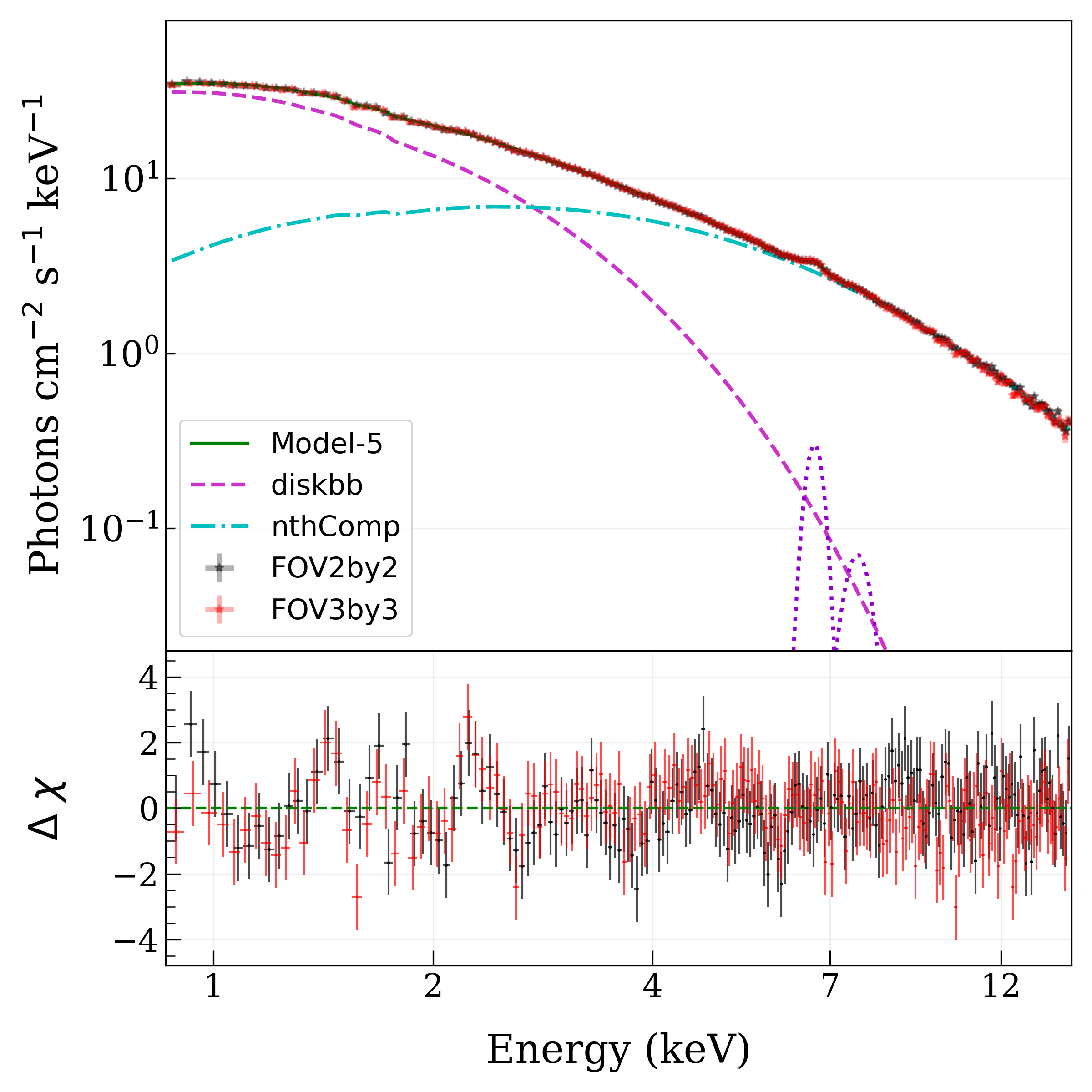}
    \includegraphics[width=0.33\linewidth]{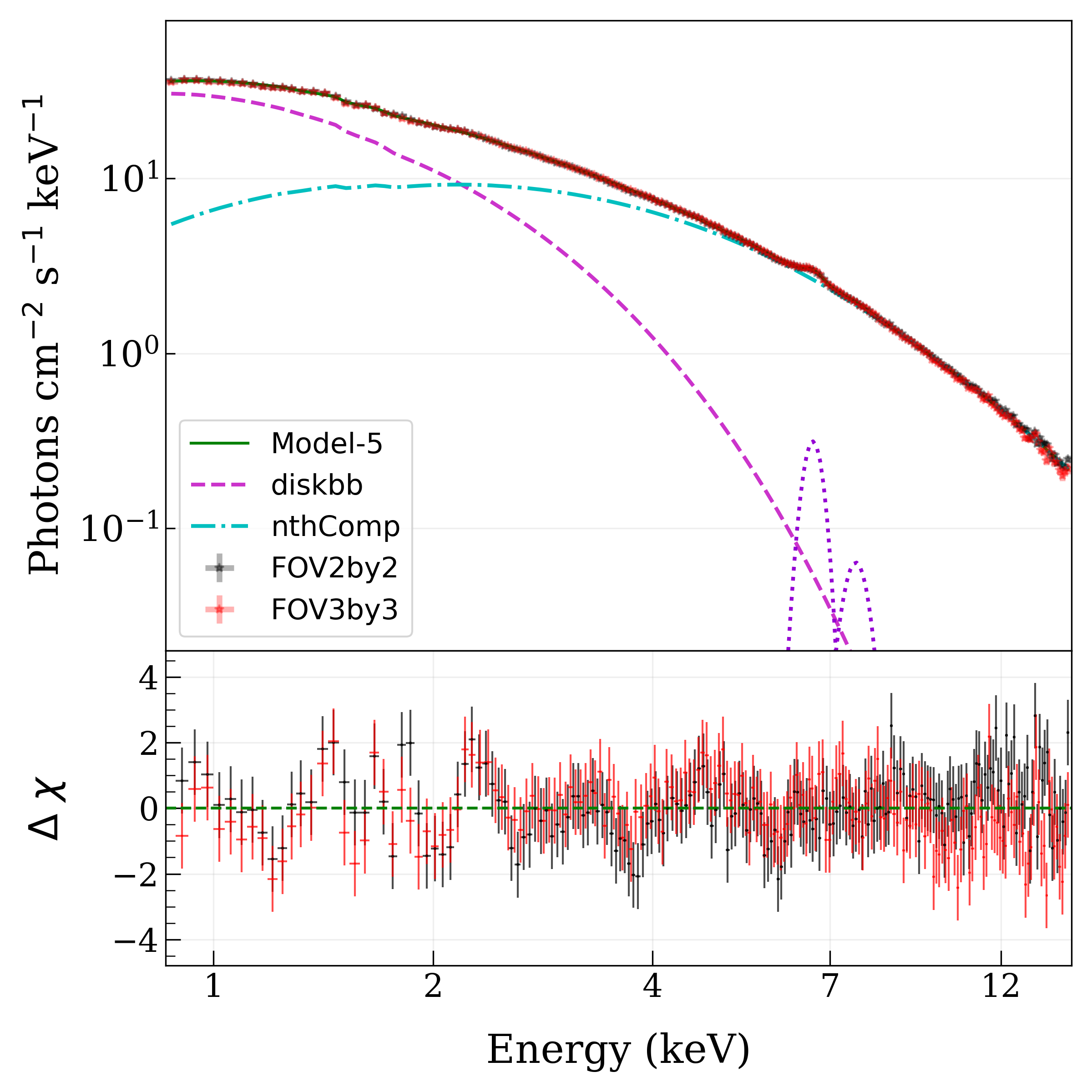}
    \includegraphics[width=0.33\linewidth]{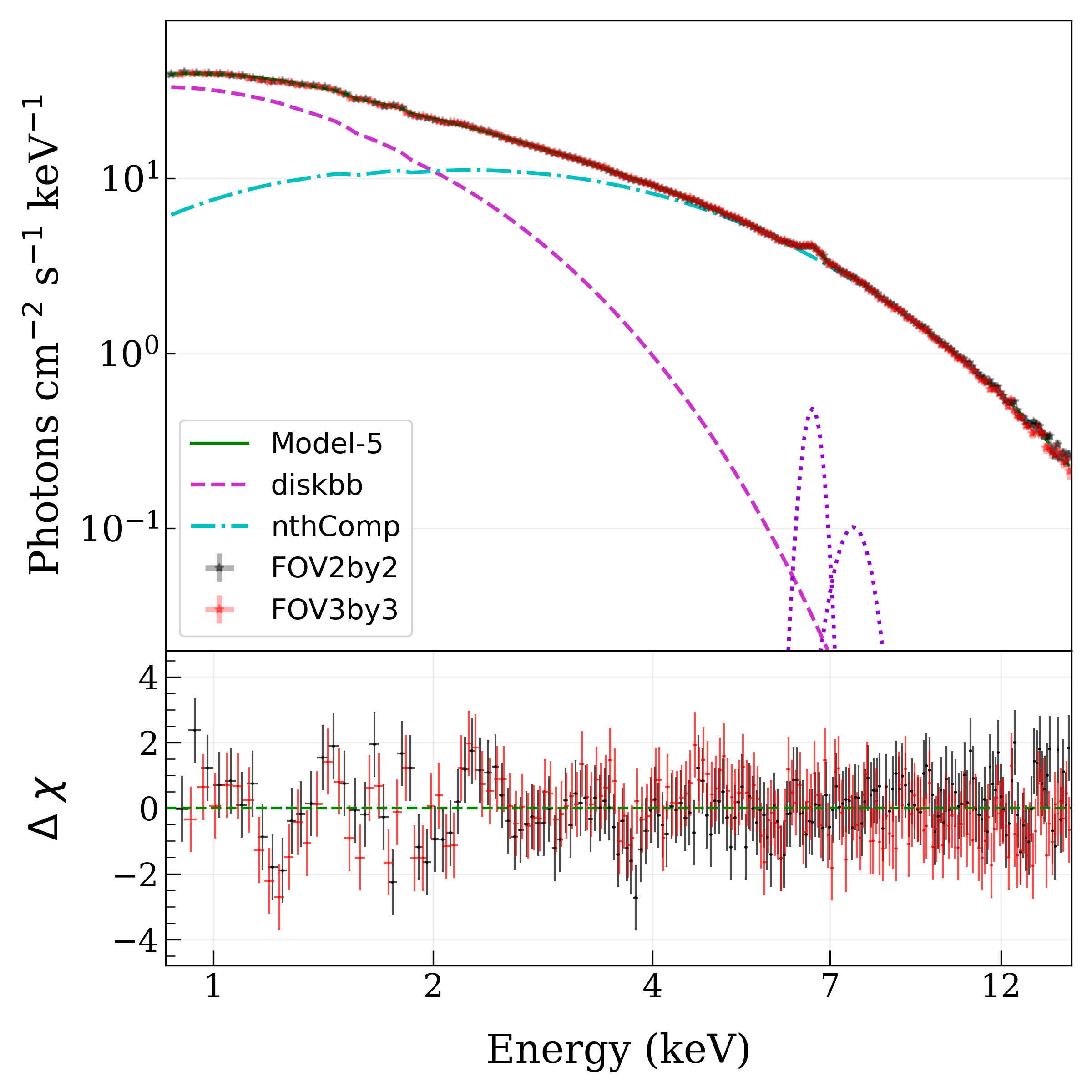}
    \caption{The 0.8-15.0} keV \textit{XSPECT} spectrum of Sco X-1 fitted using Model-5: \textit{tbabs*(diskbb+Comptb+gauss+gauss)} in the HB1 (left), NB1 (middle), and FB1 (right) in \texttt{XSPEC}. The residuals are plotted in the bottom panel. The different components in the model are also plotted along with the model (green solid line).
    \label{spec_fit_model5}
\end{figure*}

\subsubsection{Model-1}\label{subsectmodel1}

In this model, we describe the source spectra of Sco X-1 using a combination of \textit{diskbb} and \textit{nthComp} components.
The \textit{nthComp} component has electron temperature of the hot plasma ($kT_e$), the input blackbody seed photon temperature ($kT_{bb}$), the photon index ($\Gamma$) and the model normalization as free parameters.
The input seed photons for the \textit{nthComp} component was set to blackbody emission by freezing the \textit{inp\_type} to 0.
The top panel of Figure \ref{residual_plot} shows the spectral fit in the NB1 segment after modeling the continuum spectra.
We let all the parameters to vary freely during the spectral fitting.
Excess residuals are identified at $\sim$ 6.6 keV, suggesting the presence of iron K$_{\alpha}$ emission line in the spectra in all segments along the Z-track.
All the segments along the Z-track, shows the presence an additional Gaussian feature at $\sim$ 7.6 keV indicating the iron K$_{\beta}$ emission line as shown in the middle panel of Figure \ref{residual_plot}.
The two emission lines are modeled using \textit{Gaussian} components.
The improvements in the residual after the inclusion of two Gaussian features is displayed in the bottom panel.
The spectral model that provided a satisfactory fit in all segments is \textit{tbabs*(diskbb+nthComp+gauss+gauss)}.
The inner disk radius ($R_{in}$) of the system is estimated from the normalization of the \textit{diskbb} component ($N_{disc}$) using the expression;
\begin{equation*}
    N_{disc} = (R_{in}/D_{10})^2 \cos{i} ,
\end{equation*}
where $D_{10}$ is the distance to the source in units of 10 kpc and $i$ is the inclination of the source.
We consider an inclination of 44$^{\circ}$ \citep{2001ApJ...553L..27F, 2001ApJ...558..283F} and a source distance of 2.13 kpc (\citealt{2021MNRAS.502.5455A}) for the $R_{in}$ calculations. 
Spectral modeling is carried out in ten segments of the CCD and the evolution of these parameters are studied.
Table \ref{table_best_fit_model1} provide the best-fit spectral parameters obtained in all segments.
All quoted uncertainties correspond to 1$\sigma$ (68\% confidence) errors.
The unabsorbed flux values for each spectral components are estimated in the 0.5-15.0 keV energy range using \texttt{cflux} model in \texttt{XSPEC} and are reported.
The spectral fit using Model-1 for HB1, NB1 and FB1 segments are displayed along with residual in Figure \ref{spec_fit_model1}.
The different spectral components of the model are also marked.

\subsubsection{Model-2}\label{subsectmodel2}
In Model-2, the soft component in the spectra is modeled using a blackbody component (\textit{bbodyrad}) while the hard Comptonization component is modeled using \textit{nthComp} component.
The blackbody temperature ($kT_{bb}$) and the model normalization ($N_{bb}$) are the free parameters for \textit{bbodyrad} component.
The \textit{inp\_type} of the \textit{nthComp} component is now set to 1 to consider disk photons as the input seed photons.
Similar to Model-1, this model also required the addition of two Gaussian components at $\sim$6.6 keV and $\sim$7.7 keV in all the segments along the Z-track.
The $N_{H}$ is kept free during the spectral fits.
Spectral fitting is carried out in the 0.8–15.0 keV energy range using the model: \textit{tbabs*(bbodyrad+nthComp+Gauss+Gauss)} across all the segments in the CCD.
The best-fit spectral parameters along with the flux values for each components are reported in Table \ref{table_best_fit_model2}. 
The source photon radius ($R_{bb}$) is calculated using;
\begin{equation*}
    N_{bb} = (R_{bb}/D_{10})^2 ,
\end{equation*}
where $D_{10}$ is the source distance in units of 10 kpc.

\subsubsection{Model-5}\label{subsectmodel5}
This model uses a combination of \textit{diskbb}, \textit{Comptb} and \textit{Gaussian} components to explain the source spectra in the 0.8-15.0 keV energy range.
The \textit{Comptb} component has electron temperature $kT_e$, seed photon temperature $kT_s$, energy index of the Comptonization spectrum ($\alpha$), index of the seed photon spectrum ($\Gamma$), normalization of the model ($N$), and illuminating factor parameter ($logA$) as free parameters.
Here 1/(1 + A) is the fraction of the seed-photon emission directly seen by the observer and A/(1 + A) is the Comptonized emission fraction.
We fixed the $\Gamma$ to its default value of 3 and the bulk parameter $\delta$ to 0 and let all the other parameters to very freely during the spectral fit.
In this model, The best-fit value of the illuminating factor parameter $log A$ is not constrained during the spectral fit and found to be approaching the hard limit.
This indicate a high Comptonization fraction ($>$ 80\% in all segments).
In this scenario, the best-fit values of the other parameters are insensitive to any change in $log A$ \citep{2014ApJ...789...98T, 2021ApJ...913..119J, 2025PASA...42...52A}.
Hence, we fix the value of $logA$ to the maximum value of 8 in all segments.
Spectral fitting is performed in all segments and we study the evolution of best-fit spectral parameters along the Z-track.
We calculated the inner disk radius from the \textit{diskbb} normalization using the the distance and inclination values mentioned in the section \ref{subsectmodel1}.
The spectral fit using Model-5 for HB1, NB1 and FB1 segments are displayed along with residual in Figure \ref{spec_fit_model5} and the different spectral components are marked.

\begin{figure*}
    \centering
    \includegraphics[width=0.48\linewidth]{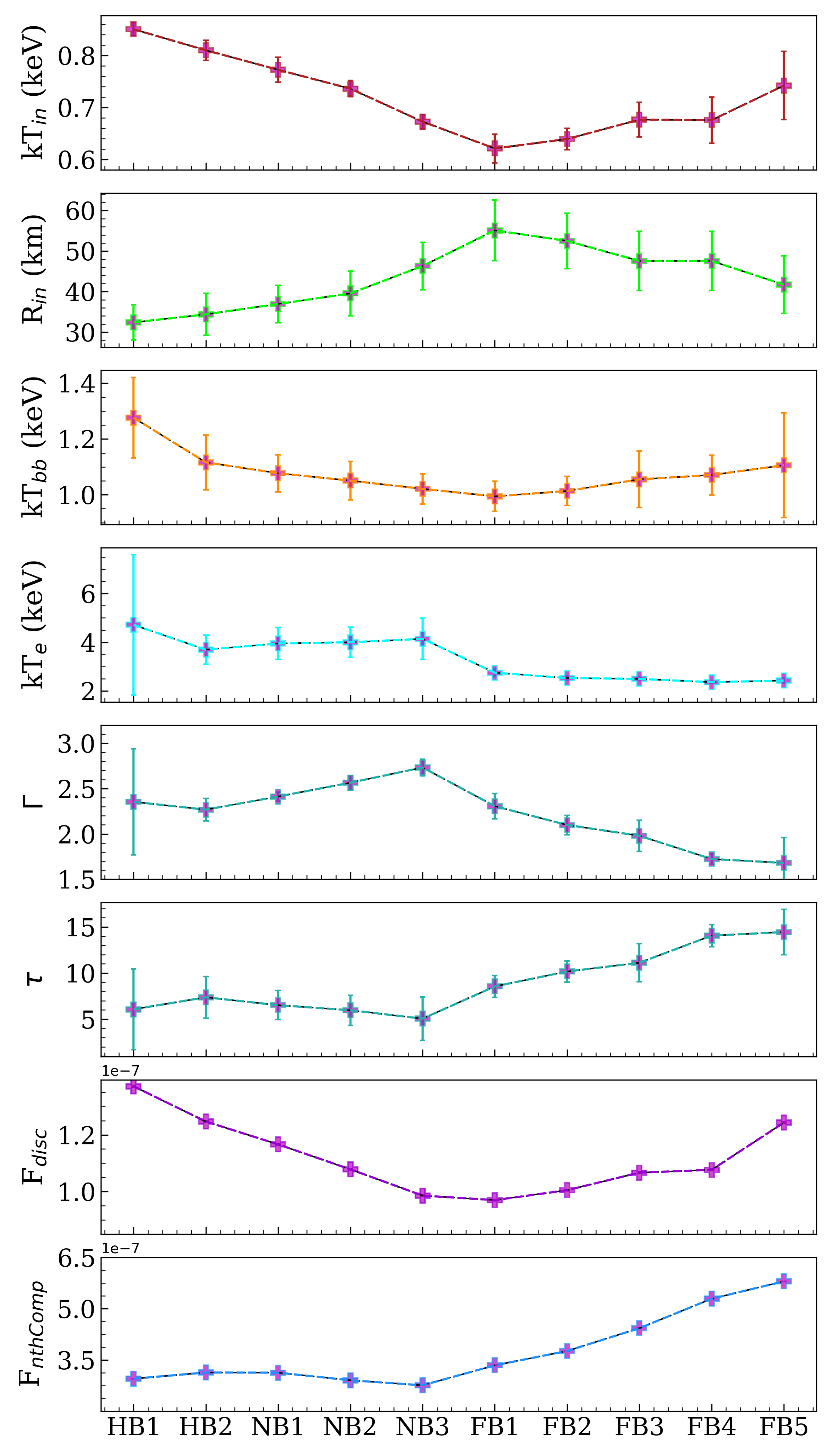}
    \includegraphics[width=0.48\linewidth]{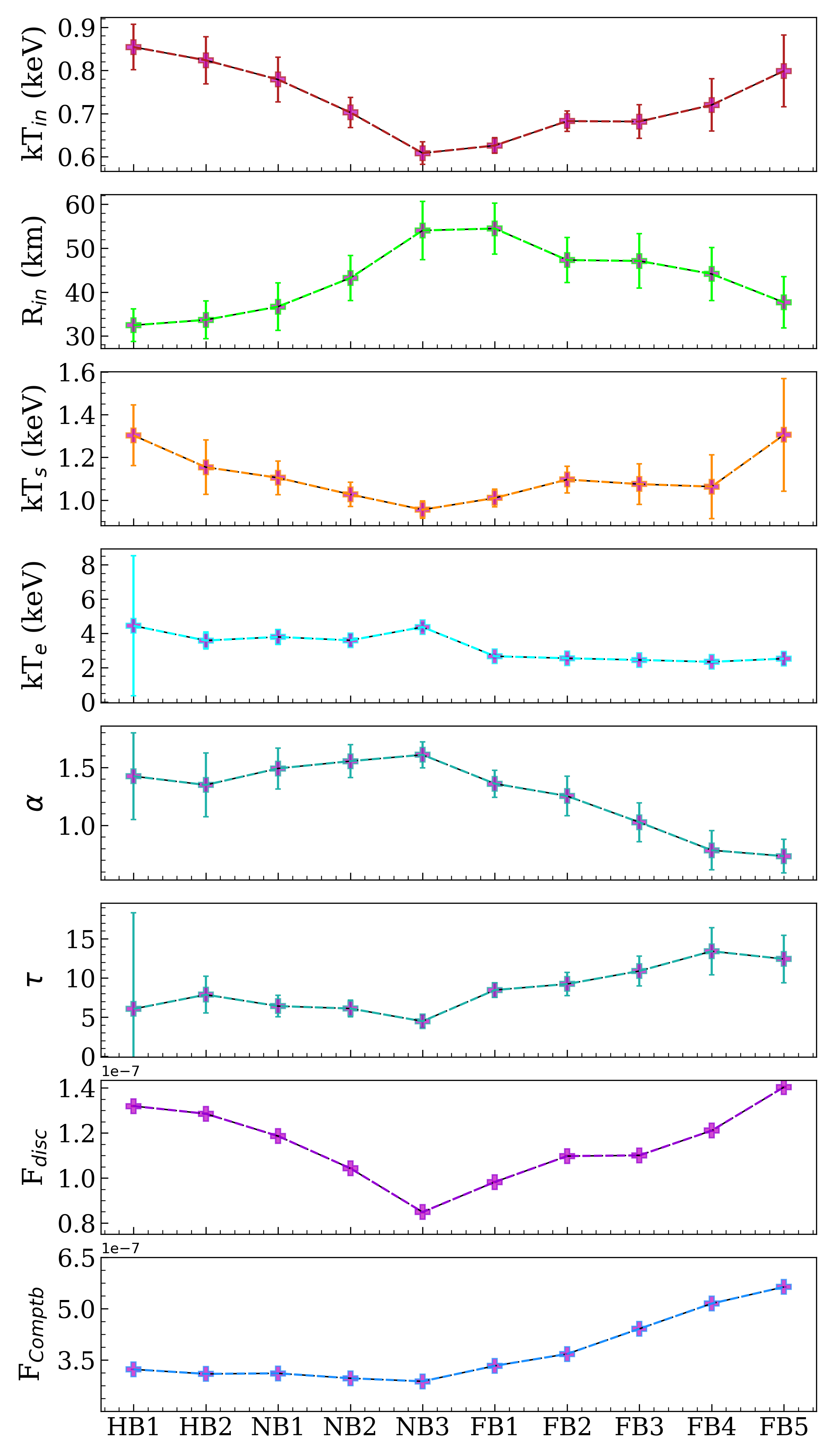}
    \caption{The evolution of best-fit spectral parameters along the Z-track segments using the models \textit{tbabs*(diskbb+nthComp+gauss+gauss)} [left], and \textit{tbabs*(diskbb+Comptb+gauss+gauss)} [right]. The error bars shown correspond to 1$\sigma$ (68\% confidence) uncertainties.}
    \label{spec_params_evol}
\end{figure*}

\subsection{Timing analysis}
For each branch in the CCD, a Power density spectrum (PDS) is created in the energy range 0.8-15.0 keV for the timing studies. 
PDS are created from 4 ms light curves (separately for FoV 2$\times$2 and FoV 3$\times$3 detectors) such that the maximum frequency in the PDS is 125 Hz (Nyquist frequency).
We calculated the rms-normalized PDS (\citealt{2002ApJ...572..392B}) for each branch in the CCD (refer Figure \ref{hid2}).
Each PDS is generated using data segments of length 2$^{13}$ bins corresponding to 32.7 sec.
The very low-frequency noise (VLFN) features in the PDS are modeled using a power law function.
PDSs are also generated separately in hard (6.0-15.0 keV) and soft (0.8-6.0 keV) energy bands to understand the energy dependence of timing features on energy. 
No QPOs are detected in any branch of the Z-track, nor in the PDS for soft- or hard-energy bands.

\section{Results}\label{res}

During the \textit{XSPECT} observations, Sco X-1 traces the complete Z-track which includes HB, NB and FB.
The right panel of Figure \ref{hid} shows the evolution of the source through the Z-track during the observations. At the beginning of observation, the source was observed to be in the upper HB and then moves close to the HA, and then the source then moves down to the NB and eventually to the FB and returned to the NB towards the end of the observation.
Strong flares are observed for a period of approximately 50 ks during the observations. 
We split the observation into ten different segments along the Z-track as shown in Figure \ref{hid2} to understand the spectral and temporal evolution of the source along the track. 
We present a comparison of the spectral fitting results obtained using different models, which are discussed in the following section.

\subsection{Spectral properties}

\subsubsection{Model-1}
We model the X-ray spectra in the 0.8-15.0 keV using a multi-color disk blackbody plus a Comptonized continuum component with seed photons supplied by single temperature blackbody emission. 
The excess residual at $\sim$ 6.6 keV required an additional \textit{Gaussian} characteristic to model the iron $K_{\alpha}$ emission.
All ten segments have shown the requirement of an additional Gaussian feature at $\sim$ 7.6 keV corresponding to the $K_\beta$ emission line of iron.
The model provided a satisfactory description of the source spectra in all the segments considered for analysis with the best-fit spectral parameter values reported in Table \ref{table_best_fit_model1}.
The left panel of Figure \ref{spec_params_evol} shows the change in the spectral components as a function of the different segments in the Z-track considered for analysis.
Both the disk and Comptonization related parameters shows significant variation along the Z-track.

It is observed that the inner radius of the accretion disk increases toward the soft apex, reaching its maximum value of approximately $\sim$55 km in the FB1 segment. Thereafter, it decreases to a lower value of approximately 42 km in FB5.
Additionally, the radius exhibits a strong negative correlation with the inner disk temperature.
The blackbody temperatures exhibit a trend in variation similar to the disk temperature (see Figure \ref{spec_params_evol}).
In HB and NB, these decrease in temperatures (for $kT_{in}$ and $kT_{bb}$) are also correlated with a decrease in the disk flux as shown in \ref{spec_params_evol}.

The source spectra are largely dominated by Comptonized emission.
The electron temperature ($kT_e$) is seen to increase along the NB towards the soft apex and then decreases along the FB from the soft apex.
In Model-1, the luminosity of the \textit{nthComp} component is approximately four times higher than the luminosity of the disk. 
We also see a large increase in the Comptonization flux during the flaring events.
Systematic variations are observed in all spectral parameters associated with Comptonized emission. 
The minimum blackbody temperature is observed at the beginning of the FB.
The spectra become softer as the source reaches the soft apex as indicated by the increase in photon index ($\Gamma$) from NB1 to NB3.
A decrease in $\Gamma$ during the flaring branch is indicative of the hardening of spectra during FB as studied by \citet{2014ApJ...789...98T}.
The increase in both Comptonization and disk flux during the flaring is seen to be a reason behind the increase in source luminosity during this period.

\begin{figure}
    \centering
    \includegraphics[width=0.88\linewidth]{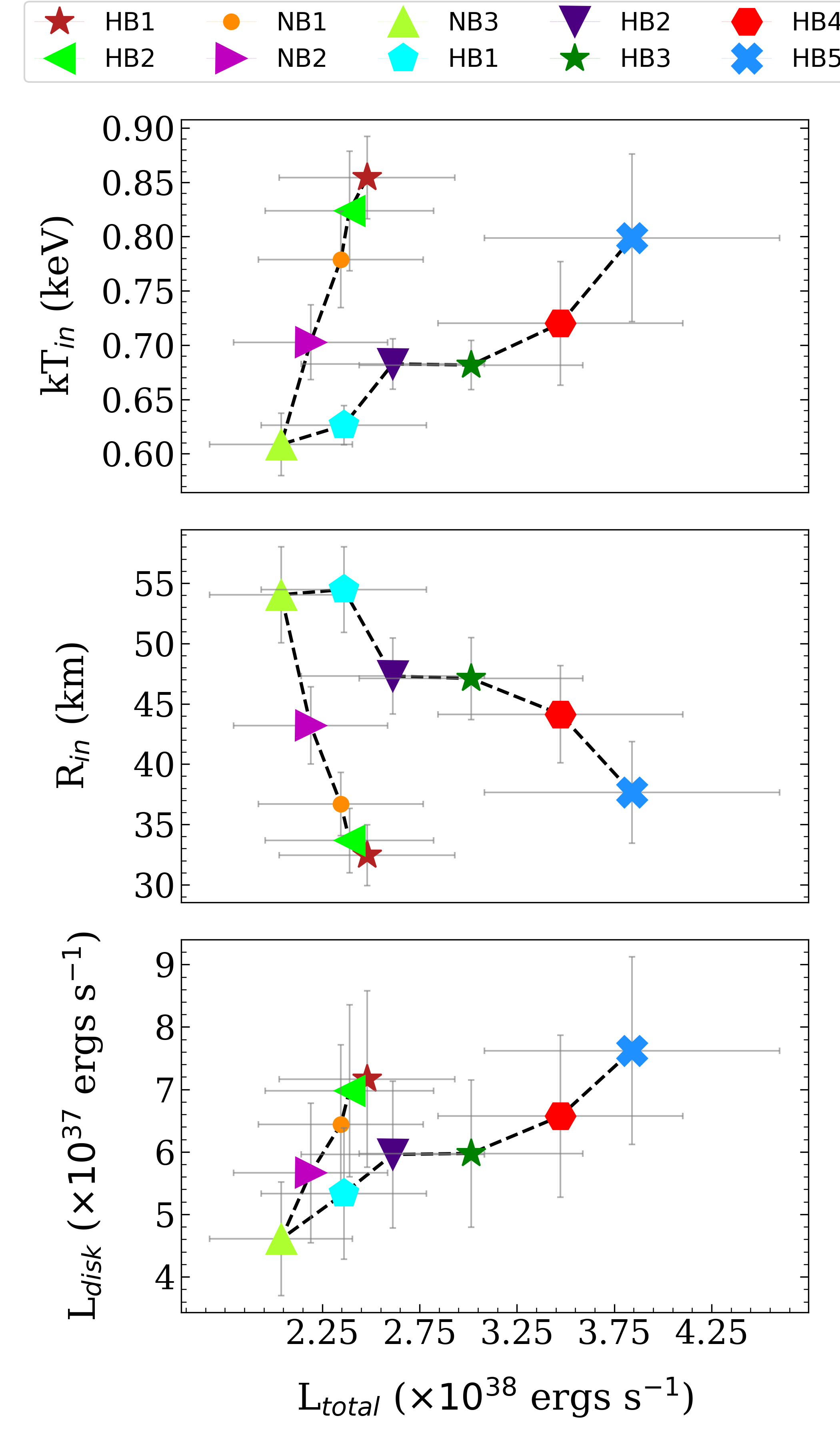}
    \caption{Evolution of the disk emission parameters (Model-5). The inner disk temperature (top) and inner disk radius (bottom) as a function of the source luminosity.}
    \label{disc_params_evol}
\end{figure}

\begin{figure}
    \centering
    \includegraphics[width=0.83\linewidth]{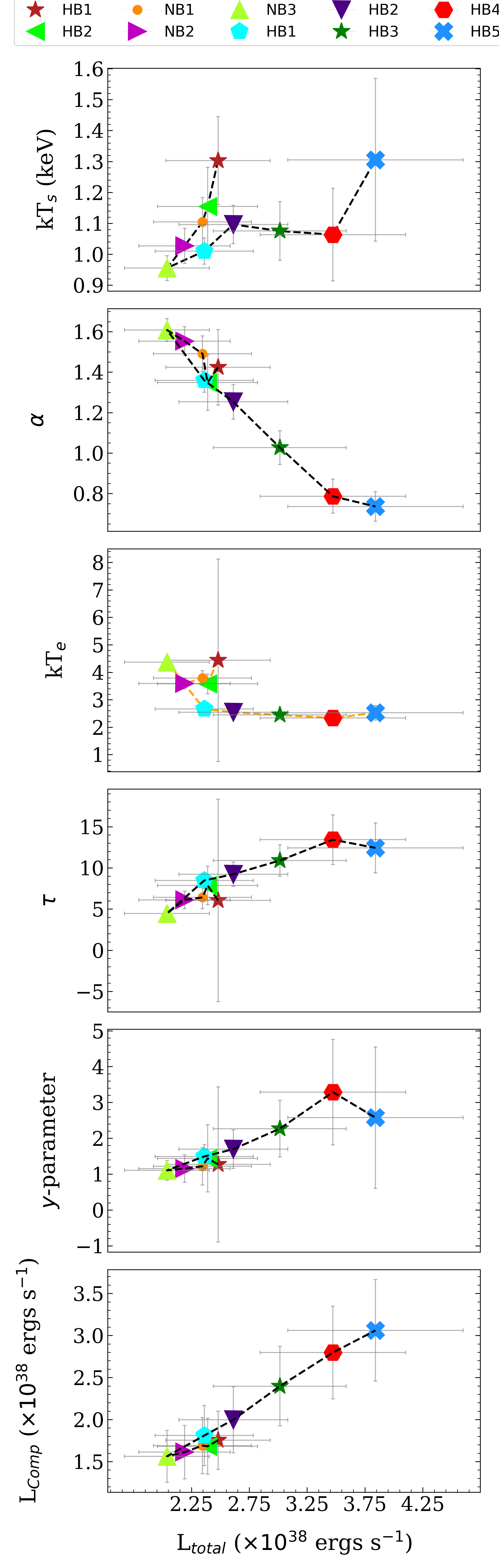}
    \caption{Evolution of the Comptonized emission (\textit{Comptb}) parameters (Model-5). The blackbody temperature, the energy Index, the electron temperature, the optical depth, $y$-parameter and and the Comptonization luminosity as a function of the total luminosity.}
    \label{nthcomp_spec_evol}
\end{figure}

The variation in the electron temperature, optical depth and the size of the equivalent seed photon radius give a comprehensive picture of the evolution of corona during the complete Z-track movement. 
We estimate the optical depth ($\tau$) of the Comptonization region from the photon index ($\Gamma$), with an assumption of spherically symmetric and uniform density corona using the expression provided by \citet{1996MNRAS.283..193Z}.
\[\Gamma = -\frac{1}{2}+\sqrt{\frac{9}{4}+\frac{1}{\frac{kT_e}{m_ec^2}(1+\frac{\tau}{3})\tau}} ,\]

where $kT_e$ is the electron temperature of the corona. 
We see that the optical depth increases slightly along the HB towards the hard apex.
However, there is decrease in $\tau$ ($\sim$ 7.5 in NB1 to $\sim$ 5 in NB3) in NB towards SA.
During the flaring events the optical depth steeply increases and reaches its maximum value of $\sim$ 14 at the end of FB (FB5).
The size of the equivalent seed photon emission ($R_W$) area is estimated by using the expression given by \citet{1999A&A...345..100I} by equalizing the bolometric luminosity of the soft photons to that of a blackbody. 
\[ R_{\mathrm{W}}=3 \times 10^{4} \times d \sqrt{\frac{f_{\mathrm{bol}}}{1+y}} /\left(k T_{\mathrm{bb}}\right)^{2}, \]

where, $d$ is the distance to the source in units of kpc, and $y$ is the Compton y-parameter (\(y=4kT_e/m_ec^2\)) and $kT_\mathrm{bb}$ is the blackbody temperature. 
The equivalent seed photon radius is found to be correlated with the inner disk radius. 
It increases from 17 km at the HB to 27 km at the end of NB and then it starts increases along FB to 20 km in FB5. 
However the error-bar on the parameter is large to derive any conclusions.

\begin{figure}
    \centering
    \includegraphics[width=0.9\linewidth]{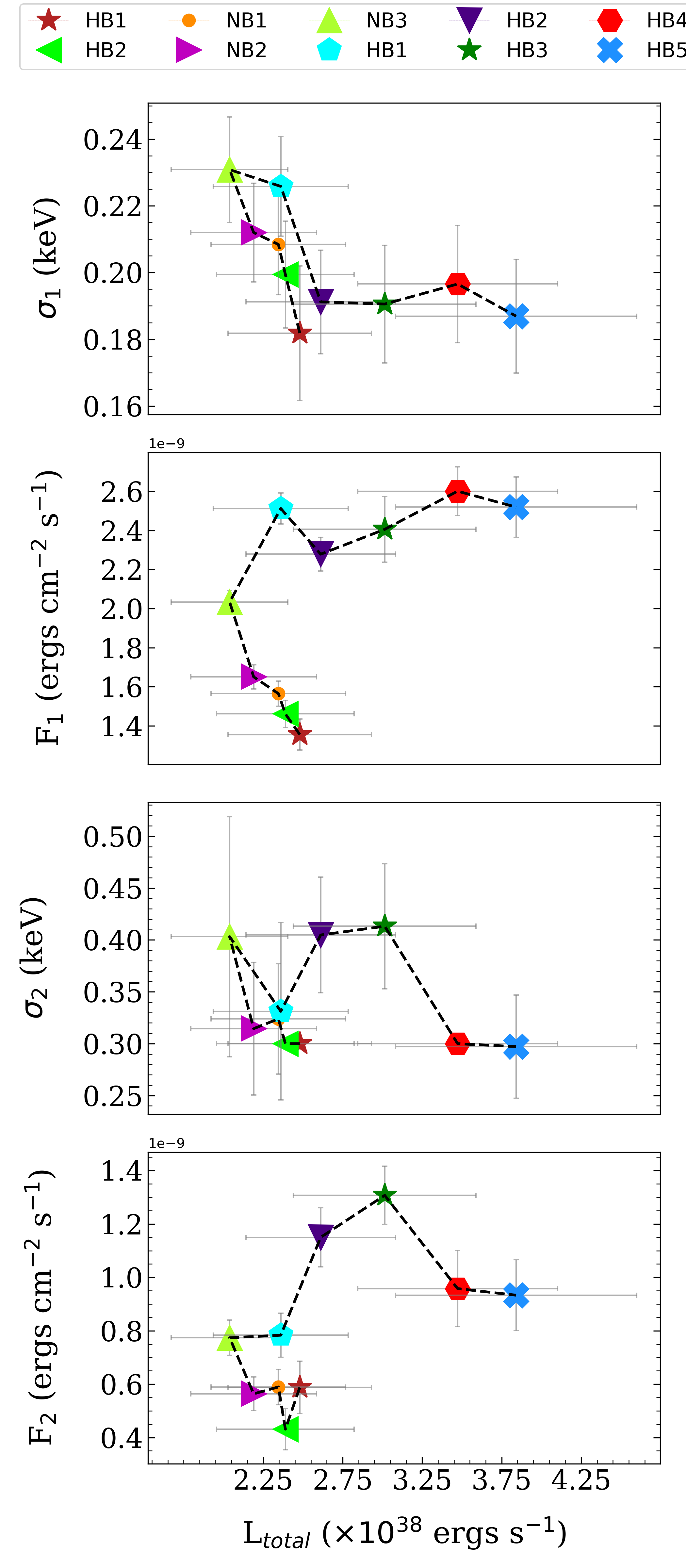}
    \caption{Evolution of the Gaussian line parameters ($\sigma$ and flux) for both $K_{\alpha}$ (subscript 1) and $K_{\beta}$ (subscript 2) from Model-5.}
    \label{gauss_spec_evol}
\end{figure}

\subsubsection{Model-2}
This model assumes that the soft emission in the spectra originates from blackbody emission, with the seed photons for Comptonization provided by the disk. 
Table \ref{table_best_fit_model2} reports the best-fit parameters from the spectral fit using Model-2.
The blackbody temperature ($kT_{bb}$) shows a variation trend similar to the inner disk temperature in Model-1.
It starts from a higher value of $\sim$1.84 at HB1 and declines to 1.39 keV at NB3 and then rises again to $\sim$1.88 keV in FB5.
This displays a significant increase in the blackbody temperature during the flaring.
In contrast to Model-1, the \textit{nthComp} parameters in this model do not exhibit a clear variation trend.
However the best-fit values of photon index ($\Gamma$) and electron temperature ($kT_e$) have different values in each brach.
The electron temperature of the corona has an average value $\sim$3.76 keV in HB,  $\sim$3.37 keV in NB and $\sim$2.57 keV in FB.
Similarly, $\Gamma$ has $\sim$1.97 in HB, $\sim$2.1 in NB and $\sim$1.78 in FB which clearly implies the hardening of the source spectra during flaring.
The blackbody source radius estimated from the normalization of the \textit{bbodyrad} model shows an increase towards the soft apex along NB and then slightly decrease along the FB correlated with the inner disk radius in Model-1.

Comptonized emission dominates the spectra in all segments for the model. 
A substantial increase in the Comptonized flux evident from NB3 to FB5.

\begin{landscape}
\begin{table}
\renewcommand{\arraystretch}{1.4}
    \caption{The best-fit parameters from spectral fitting using Model-1. The free parameter for the \textit{diskbb} component are the inner disk temperature ($kT_{in}$), and the model normalization ($N_{disk}$). The photon index ($\Gamma$), the electron temperature ($kT_e$), the blackbody seed photon temperature ($kT_{bb}$) and the model normalization ($N_{nthComp}$) are the free parameters for the \textit{nthComp} component. The line energy ($E$), width ($\sigma$) and the normalizations ($N_{gauss}$) for the two separate Gaussian components are reported with subscript 1 and 2 for Gaussian-1 and 2 respectively. All quoted uncertainties correspond to 1$\sigma$ (68\% confidence) errors.} 
    \label{table_best_fit_model1}
    \begin{tabular}{llcccccccccc}
        \hline
        \multicolumn{9}{c}{Model-1: \texttt{const*tbabs*(diskbb+nthComp+Gauss+Gauss)}}\\
        \hline
        Model & Parameter & HB1 & HB2 & NB1 & NB2 & NB3 & FB1 & FB2 & FB3 & FB4 & FB5\\
        \hline
        \texttt{Const} & 2$\times$2 & 1 & 1 & 1 & 1 & 1 & 1 & 1 & 1 & 1 & 1 \\ 

        & 3$\times$3 & 1.091$\pm$0.001 & 1.091$\pm$0.001 & 1.090$\pm$0.002 & 1.095$\pm$0.001 &  1.099$\pm$0.001 & 1.093$\pm$0.001 & 1.108$\pm$0.001 & 1.091$\pm$0.001 & 1.091$\pm$0.01& 1.092$\pm$0.01\\ 

        \texttt{tbabs} & $N_{H} (\times 10^{22} \text{cm}^{-2}) $ & 0.14$^{*}$ &0.14$_{-0.01}^{+0.01}$ &0.14$_{-0.01}^{+0.01}$ &0.13$_{-0.01}^{+0.01}$ &0.14$^{*}$ &0.14$_{-0.0}^{+0.0}$ &0.14$_{-0.0}^{+0.0}$ &0.13$_{-0.01}^{+0.01}$ &0.14$^{*}$ &0.14$_{-0.01}^{+0.01}$ \\ 
        
        \texttt{diskbb} & $kT_{\rm in}$ (keV) & 0.85$_{-0.04}^{+0.01}$ &0.81$_{-0.05}^{+0.02}$ &0.77$_{-0.05}^{+0.02}$ &0.74$_{-0.03}^{+0.02}$ &0.67$_{-0.02}^{+0.01}$ &0.62$_{-0.03}^{+0.03}$ &0.64$_{-0.02}^{+0.02}$ &0.68$_{-0.05}^{+0.03}$ &0.68$_{-0.03}^{+0.04}$ &0.74$_{-0.07}^{+0.07}$ \\
        
        & $N_{\rm disk}$ ($\times 10^4$) & 1.6$_{-0.1}^{+0.2}$ &1.8$_{-0.1}^{+0.3}$ &2.1$_{-0.1}^{+0.1}$ &2.4$_{-0.1}^{+0.3}$ &3.3$_{-0.2}^{+0.2}$ &4.7$_{-0.7}^{+0.6}$ &4.2$_{-0.4}^{+0.4}$ &3.5$_{-0.6}^{+0.7}$ &3.7$_{-0.8}^{+0.4}$ &2.7$_{-0.8}^{+0.6}$ \\
        
        & $R_{in}$ (km) &  32.0$_{-4.0}^{+4.0}$ &34.0$_{-4.0}^{+5.0}$ &37.0$_{-5.0}^{+5.0}$ &40.0$_{-5.0}^{+6.0}$ &46.0$_{-6.0}^{+6.0}$ &55.0$_{-8.0}^{+8.0}$ &53.0$_{-7.0}^{+7.0}$ &48.0$_{-7.0}^{+7.0}$ &48.0$_{-7.0}^{+7.0}$ &42.0$_{-8.0}^{+7.0}$ \\
        
        \texttt{nthComp} & $\Gamma$ & 2.35$_{-0.58}^{+0.27}$ &2.27$_{-0.12}^{+0.12}$ &2.41$_{-0.05}^{+0.1}$ &2.56$_{-0.08}^{+0.11}$ &2.73$_{-0.09}^{+0.08}$ &2.31$_{-0.14}^{+0.11}$ &2.1$_{-0.11}^{+0.08}$ &1.98$_{-0.17}^{+0.09}$ &1.72$_{-0.06}^{+0.06}$ &1.68$_{-0.28}^{+0.09}$ \\
        
        & $kT_e$ (keV) & 4.72$_{-4.14}^{+1.61}$ &3.7$_{-0.9}^{+0.29}$ &3.95$_{-0.8}^{+0.5}$ &4.0$_{-0.9}^{+0.34}$ &4.14$_{-1.43}^{+0.26}$ &2.75$_{-0.18}^{+0.13}$ &2.53$_{-0.1}^{+0.08}$ &2.5$_{-0.07}^{+0.07}$ &2.36$_{-0.04}^{+0.04}$ &2.43$_{-0.17}^{+0.05}$ \\
        
        & $kT_{\rm bb}$ (keV) & 1.28$_{-0.14}^{+0.14}$ &1.12$_{-0.1}^{+0.1}$ &1.08$_{-0.07}^{+0.06}$ &1.05$_{-0.07}^{+0.05}$ &1.02$_{-0.05}^{+0.02}$ &0.99$_{-0.05}^{+0.05}$ &1.01$_{-0.05}^{+0.05}$ &1.06$_{-0.1}^{+0.07}$ &1.07$_{-0.07}^{+0.09}$ &1.11$_{-0.19}^{+0.15}$ \\
        
        & $N_{\rm nthComp}$ & 5.21$_{-1.13}^{+0.72}$ &7.23$_{-1.48}^{+1.4}$ &8.19$_{-0.53}^{+1.16}$ &8.48$_{-0.79}^{+1.01}$ &9.03$_{-0.45}^{+0.77}$ &10.55$_{-0.96}^{+0.88}$ &10.56$_{-0.8}^{+0.74}$ &10.78$_{-1.09}^{+1.34}$ &12.57$_{-1.83}^{+1.17}$ &10.4$_{-2.31}^{+2.05}$ \\

        \texttt{Gaussian-1} & $E_1$ (keV) & 6.69$_{-0.02}^{+0.02}$ &6.69$_{-0.01}^{+0.01}$ &6.71$_{-0.01}^{+0.01}$ &6.7$_{-0.01}^{+0.01}$ &6.71$_{-0.01}^{+0.02}$ &6.7$_{-0.01}^{+0.02}$ &6.68$_{-0.01}^{+0.01}$ &6.7$_{-0.01}^{+0.01}$ &6.68$_{-0.02}^{+0.02}$ &6.69$_{-0.02}^{+0.02}$ \\
        
        & $\sigma_1$ (keV) & 0.18$_{-0.02}^{+0.02}$ &0.2$_{-0.02}^{+0.02}$ &0.21$_{-0.01}^{+0.01}$ &0.21$_{-0.01}^{+0.01}$ &0.23$_{-0.02}^{+0.02}$ &0.23$_{-0.01}^{+0.02}$ &0.19$_{-0.02}^{+0.02}$ &0.19$_{-0.02}^{+0.02}$ &0.19$_{-0.02}^{+0.02}$ &0.19$_{-0.02}^{+0.02}$ \\
        
        & $N_{\rm gauss}$ & 0.13$_{-0.01}^{+0.01}$ &0.14$_{-0.01}^{+0.01}$ &0.15$_{-0.01}^{+0.01}$ &0.15$_{-0.01}^{+0.01}$ &0.19$_{-0.01}^{+0.01}$ &0.23$_{-0.01}^{+0.02}$ &0.22$_{-0.02}^{+0.02}$ 
        &0.22$_{-0.02}^{+0.02}$ &0.23$_{-0.02}^{+0.02}$ &0.24$_{-0.02}^{+0.02}$ \\

        \texttt{Gaussian-2} & $E_2$ (keV) & 7.66$_{-0.07}^{+0.08}$ &7.72$_{-0.09}^{+0.1}$ &7.69$_{-0.05}^{+0.05}$ &7.66$_{-0.06}^{+0.07}$ &7.65$_{-0.07}^{+0.12}$ &7.67$_{-0.06}^{+0.11}$ &7.59$_{-0.07}^{+0.11}$ &7.5$_{-0.14}^{+0.15}$ &7.54$_{-0.12}^{+0.13}$ &7.64$_{-0.07}^{+0.09}$ \\
        
        & $\sigma_2$ (keV) & 0.3$_{-0.0}^{+0.0}$ &0.3$_{-0.0}^{+0.0}$ &0.3$_{-0.0}^{+0.0}$ &0.3$_{-0.0}^{+0.0}$ &0.34$_{-0.13}^{+0.07}$ &0.33$_{-0.18}^{+0.07}$ &0.42$_{-0.12}^{+0.08}$ &0.42$_{-0.11}^{+0.13}$ &0.48$_{-0.11}^{+0.11}$ &0.3$_{-0.09}^{+0.06}$ \\
        
        & $N_{\rm gauss}$ & 0.05$_{-0.01}^{+0.01}$ &0.04$_{-0.01}^{+0.01}$ &0.05$_{-0.01}^{+0.01}$ &0.04$_{-0.01}^{+0.01}$ &0.05$_{-0.02}^{+0.01}$ &0.06$_{-0.03}^{+0.01}$ &0.1$_{-0.03}^{+0.02}$ 
        &0.11$_{-0.03}^{+0.03}$ &0.12$_{-0.03}^{+0.03}$ &0.08$_{-0.02}^{+0.02}$ \\

        $\chi^2/dof$  & &  376/362 & 427/376 & 385/386 & 383/378 & 349/376 & 349/367 & 311/369 & 336/366 & 352/368 & 381/356 \\ 
        \hline
        
        & $\tau$   &  6.07$\pm$4.38 & 7.37$\pm$2.25 & 6.52$\pm$1.58 & 5.97$\pm$1.63 & 5.06$\pm$2.35 & 8.56$\pm$1.17 & 10.17$\pm$1.15 & 11.12$\pm$2.07 & 14.06$\pm$1.2 & 14.46$\pm$2.45 \\

        & $y$  & 1.36$\pm$1.96 & 1.57$\pm$0.96 & 1.32$\pm$0.64 & 1.12$\pm$0.61 & 0.94$\pm$0.81 & 1.58$\pm$0.43 & 2.05$\pm$0.46 & 2.42$\pm$0.9 & 3.66$\pm$0.63 & 3.96$\pm$1.34 \\
        
        & $R_{W}$ (km)   &  17.0$_{-11.0}^{+11.0}$ &21.0$_{-9.0}^{+9.0}$ &24.0$_{-9.0}^{+9.0}$ &25.0$_{-10.0}^{+10.0}$ &27.0$_{-13.0}^{+13.0}$ &27.0$_{-8.0}^{+8.0}$ &1384.0$_{-796.0}^{+789.0}$ &23.0$_{-9.0}^{+8.0}$ &21.0$_{-7.0}^{+6.0}$ &20.0$_{-9.0}^{+8.0}$ \\

        & $F_{disk}$ ($\times 10^{-7}$)  & 1.371$_{-0.002}^{+0.002}$ &1.247$_{-0.002}^{+0.002}$ &1.166$_{-0.002}^{+0.002}$ &1.078$_{-0.002}^{+0.002}$ &0.986$_{-0.002}^{+0.002}$ &0.97$_{-0.002}^{+0.002}$ &1.005$_{-0.002}^{+0.002}$ &1.067$_{-0.003}^{+0.003}$ &1.077$_{-0.003}^{+0.003}$ &1.243$_{-0.003}^{+0.003}$ \\
        
        & $F_{Comp}$ ($\times 10^{-7}$)  & 2.954$_{-0.004}^{+0.004}$ &3.135$_{-0.003}^{+0.003}$ &3.131$_{-0.003}^{+0.003}$ &2.909$_{-0.003}^{+0.003}$ &2.762$_{-0.003}^{+0.003}$ &3.345$_{-0.003}^{+0.003}$ &3.763$_{-0.004}^{+0.004}$ &4.428$_{-0.005}^{+0.005}$ &5.288$_{-0.005}^{+0.005}$ &5.795$_{-0.006}^{+0.006}$ \\
        
        & $F_{gauss}$ ($\times 10^{-9}$)  &  0.59$_{-0.1}^{+0.1}$ &0.44$_{-0.08}^{+0.08}$ &0.6$_{-0.07}^{+0.07}$ &0.55$_{-0.06}^{+0.06}$ &0.61$_{-0.06}^{+0.06}$ &0.79$_{-0.08}^{+0.08}$ &1.23$_{-0.11}^{+0.11}$ &2.41$_{-0.11}^{+0.11}$ &2.44$_{-0.12}^{+0.12}$ &2.58$_{-0.13}^{+0.13}$ \\

        & $L_{\text{total}}$ ($\times 10^{38}$)  & 2.36$_{-0.49}^{+0.43}$ &2.39$_{-0.5}^{+0.43}$ &2.35$_{-0.49}^{+0.42}$ &2.18$_{-0.45}^{+0.39}$ &2.05$_{-0.42}^{+0.37}$ &2.36$_{-0.49}^{+0.42}$ &8198.29$_{-4431.46}^{+1477.58}$ &3.0$_{-0.65}^{+0.57}$ &3.48$_{-0.75}^{+0.66}$ &3.84$_{-0.85}^{+0.75}$ \\
        \hline 
    \end{tabular}
    \\
        $^*$ represents frozen parameters.\\
        $R_{in}$ is calculated assuming a source inclination of 46$^{\circ}$ and a distance of 2.13$^{+0.26}_{-0.21}$ kpc. \\
        Flux values in $\mathrm{erg\,cm^{-2}\,s^{-1}}$ are calculated in the 0.5-15.0 keV energy range using \texttt{cflux} command in \texttt{XSPEC}.\\
        Total luminosity is given in ergs s$^{-1}$ unit.\\
\end{table}
\end{landscape}

\begin{landscape}
\begin{table}
\renewcommand{\arraystretch}{1.4}
    \caption{The best-fit parameters from spectral fitting using Model-2. The free parameter for the \textit{bbodyrad} component are the blackbody temperature ($kT_{bb}$), and the model normalization ($N_{bb}$). The photon index ($\Gamma$), the electron temperature ($kT_e$), the seed photon temperature ($kT_{s}$) and the model normalization ($N_{nthComp}$) are the free parameters for the \textit{nthComp} component. The line energy ($E$), width ($\sigma$) and the normalizations ($N_{gauss}$) for the two separate Gaussian components are reported with subscript 1 and 2 for Gaussian-1 and 2 respectively. All quoted uncertainties correspond to 1$\sigma$ (68\% confidence) errors.} 
    \label{table_best_fit_model2}
    \begin{tabular}{llcccccccccc}
        \hline
        \multicolumn{9}{c}{Model-2: \texttt{const*tbabs*(bbodyrad+nthComp+Gauss+Gauss)}}\\
        \hline
        Model & Parameter & HB1 & HB2 & NB1 & NB2 & NB3 & FB1 & FB2 & FB3 & FB4 & FB5\\
        \hline
        \texttt{Const} & 2$\times$2 & 1 & 1 & 1 & 1 & 1 & 1 & 1 & 1 & 1 & 1 \\ 

        & 3$\times$3 & 1.091$\pm$0.001 & 1.091$\pm$0.001 & 1.090$\pm$0.002 & 1.095$\pm$0.001 &  1.099$\pm$0.001 & 1.093$\pm$0.001 & 1.108$\pm$0.001 & 1.091$\pm$0.001 & 1.091$\pm$0.01& 1.092$\pm$0.01\\ 

        \texttt{tbabs} & $N_{H} (\times 10^{22} \text{cm}^{-2}) $ & 0.14$_{+0.01}^{-0.01}$ &0.15$_{+0.01}^{-0.01}$ &0.15$_{+0.0}^{-0.01}$ &0.14$_{+0.01}^{-0.01}$ &0.13$_{+0.01}^{-0.01}$ &0.14$_{+0.0}^{-0.0}$ &0.13$_{+0.01}^{-0.01}$ &0.13$_{+0.01}^{-0.01}$ &0.15$_{+0.01}^{-0.01}$ &0.14$_{+0.01}^{-0.01}$ \\ 
        
        \texttt{bbodyrad} & $kT_{\rm bb}$ (keV) & 1.84$_{-0.07}^{+0.06}$ &1.48$_{-0.14}^{+0.18}$ &1.35$_{-0.17}^{+0.07}$ &1.32$_{-0.06}^{+0.09}$ &1.39$_{-0.06}^{+0.05}$ &1.36$_{-0.05}^{+0.05}$ &1.55$_{-0.05}^{+0.06}$ &1.55$_{-0.07}^{+0.1}$ &1.53$_{-0.17}^{+0.17}$ &1.88$_{-0.02}^{+0.08}$ \\
        
        & $N_{\rm bb}$ ($\times 10^3$) & 0.44$_{-0.05}^{+0.05}$ &0.83$_{-0.22}^{+0.14}$ &1.49$_{-0.19}^{+0.44}$ &1.87$_{-0.77}^{+0.25}$ &1.95$_{-0.11}^{+0.2}$ &2.38$_{-0.16}^{+0.14}$ &2.09$_{-0.11}^{+0.13}$ &2.02$_{-0.17}^{+0.18}$ &1.36$_{-0.41}^{+0.22}$ &2.12$_{-0.24}^{+0.53}$ \\
        
        & $R_{bb}$ (km) &  4.49$_{-0.6}^{+0.5}$ &6.13$_{-1.11}^{+0.8}$ &8.21$_{-1.14}^{+1.46}$ &9.21$_{-2.19}^{+1.1}$ &9.41$_{-1.18}^{+1.04}$ &10.4$_{-1.31}^{+1.07}$ &9.75$_{-1.22}^{+1.01}$ &9.57$_{-1.24}^{+1.03}$ &7.86$_{-1.51}^{+1.0}$ &9.82$_{-1.32}^{+1.56}$ \\
        
        \texttt{nthComp} & $\Gamma$ & 1.99$_{-0.02}^{+0.02}$ &1.96$_{-0.06}^{+0.07}$ &1.96$_{-0.11}^{+0.04}$ &2.01$_{-0.05}^{+0.07}$ &2.17$_{-0.08}^{+0.08}$ &1.88$_{-0.07}^{+0.05}$ &1.95$_{-0.13}^{+0.1}$ &1.76$_{-0.1}^{+0.08}$ &1.6$_{-0.13}^{+0.05}$ &1.69$_{-0.4}^{+0.23}$ \\
        
        & $kT_e$ (keV) & 4.15$_{-0.0}^{+0.0}$ &3.37$_{-0.38}^{+0.24}$ &3.28$_{-0.44}^{+0.13}$ &3.22$_{-0.18}^{+0.19}$ &3.61$_{-0.69}^{+0.33}$ &2.68$_{-0.19}^{+0.12}$ &2.9$_{-0.42}^{+0.25}$ &2.61$_{-0.23}^{+0.17}$ &2.39$_{-0.25}^{+0.08}$ &2.27$_{-3.68}^{+0.64}$ \\
        
        & $kT_{\rm s}$ (keV) & 0.74$_{-0.02}^{+0.02}$ &0.77$_{-0.05}^{+0.04}$ &0.73$_{-0.08}^{+0.04}$ &0.69$_{-0.04}^{+0.07}$ &0.69$_{-0.04}^{+0.04}$ &0.51$_{-0.03}^{+0.03}$ &0.61$_{-0.05}^{+0.05}$ &0.57$_{-0.04}^{+0.05}$ &0.53$_{-0.04}^{+0.05}$ &0.7$_{-0.08}^{+0.09}$ \\
        
        & $N_{\rm nthComp}$ & 45.43$_{-0.02}^{+0.51}$ &45.76$_{-0.05}^{+0.48}$ &46.08$_{-0.08}^{+0.46}$ &44.99$_{-0.04}^{+0.51}$ &44.66$_{-0.04}^{+0.54}$ &49.63$_{-0.03}^{+0.22}$ &48.15$_{-0.05}^{+0.86}$ &50.14$_{-0.04}^{+0.93}$ &53.91$_{-0.04}^{+1.75}$ &50.18$_{-0.08}^{+0.72}$ \\

        \texttt{Gaussian-1} & $E_1$ (keV) & 6.68$_{-0.02}^{+0.02}$ &6.67$_{-0.01}^{+0.01}$ &6.69$_{-0.01}^{+0.01}$ &6.68$_{-0.01}^{+0.01}$ &6.71$_{-0.01}^{+0.01}$ &6.7$_{-0.01}^{+0.01}$ &6.68$_{-0.01}^{+0.01}$ &6.72$_{-0.01}^{+0.01}$ &6.7$_{-0.01}^{+0.02}$ &6.69$_{-0.02}^{+0.02}$ \\
        
        & $\sigma_1$ (keV) & 0.19$_{-0.02}^{+0.02}$ &0.19$_{-0.02}^{+0.02}$ &0.2$_{-0.02}^{+0.02}$ &0.2$_{-0.02}^{+0.01}$ &0.23$_{-0.01}^{+0.01}$ &0.22$_{-0.01}^{+0.01}$ &0.19$_{-0.01}^{+0.01}$ &0.2$_{-0.02}^{+0.02}$ &0.2$_{-0.02}^{+0.02}$ &0.19$_{-0.02}^{+0.02}$ \\
        
        & $N_{\rm gauss}$ & 0.13$_{-0.01}^{+0.01}$ &0.13$_{-0.01}^{+0.01}$ &0.14$_{-0.01}^{+0.01}$ &0.14$_{-0.01}^{+0.01}$ &0.18$_{-0.01}^{+0.01}$ &0.23$_{-0.01}^{+0.01}$ &0.21$_{-0.01}^{+0.01}$ &0.06$_{-0.01}^{+0.01}$ &0.24$_{-0.02}^{+0.02}$ &0.23$_{-0.02}^{+0.02}$ \\

        \texttt{Gaussian-2} & $E_2$ (keV) & 7.65$_{-0.07}^{+0.1}$ &7.7$_{-0.14}^{+0.16}$ &7.64$_{-0.09}^{+0.11}$ &7.58$_{-0.11}^{+0.12}$ &7.65$_{-0.07}^{+0.1}$ &7.68$_{-0.06}^{+0.08}$ &7.6$_{-0.07}^{+0.1}$ &7.58$_{-0.1}^{+0.11}$ &7.62$_{-0.08}^{+0.09}$ &7.64$_{-0.08}^{+0.09}$ \\
        
        & $\sigma_2$ (keV) &0.33$_{-0.21}^{+0.09}$ &0.32$_{-0.16}^{+0.14}$ &0.36$_{-0.12}^{+0.14}$ &0.38$_{-0.13}^{+0.11}$ &0.31$_{-0.1}^{+0.06}$ &0.3$_{-0.11}^{+0.07}$ &0.36$_{-0.1}^{+0.07}$ &0.3$_{-0.0}^{+0.0}$ &0.3$_{-0.0}^{+0.0}$ &0.28$_{-0.09}^{+0.06}$ \\
        
        & $N_{\rm gauss}$ ($\times 10^{-2}$) & 0.6$_{-0.3}^{+0.1}$ &0.7$_{-0.2}^{+0.2}$ &0.8$_{-0.2}^{+0.3}$ &0.8$_{-0.3}^{+0.2}$ &0.4$_{-0.1}^{+0.1}$ &0.5$_{-0.2}^{+0.1}$ &0.8$_{-0.2}^{+0.1}$ &11.0$_{-0.0}^{+0.0}$ &0.8$_{-0.1}^{+0.1}$ &0.7$_{-0.2}^{+0.2}$ \\

        $\chi^2/dof$  & &  396/361 & 437/372 & 391/383 & 402/373 & 381/367 & 379/368 & 337/367 & 367/366 & 369/368 & 395/356 \\ 
        \hline

        & $F_{bbodyrad}$ ($\times 10^{-7}$)  & 0.53$_{-0.01}^{+0.01}$ &0.42$_{-0.01}^{+0.01}$ &0.52$_{-0.01}^{+0.01}$ &0.6$_{-0.01}^{+0.01}$ &0.77$_{-0.0}^{+0.0}$ &0.91$_{-0.01}^{+0.01}$ &1.26$_{-0.01}^{+0.01}$ &1.22$_{-0.01}^{+0.01}$ &0.83$_{-0.01}^{+0.01}$ &2.74$_{-0.01}^{+0.01}$ \\
        
        & $F_{Comp}$ ($\times 10^{-7}$)  & 3.8$_{-0.01}^{+0.01}$ &3.96$_{-0.01}^{+0.01}$ &3.78$_{-0.01}^{+0.01}$ &3.39$_{-0.01}^{+0.01}$ &2.96$_{-0.01}^{+0.01}$ &3.41$_{-0.01}^{+0.01}$ &3.49$_{-0.01}^{+0.01}$ &4.29$_{-0.01}^{+0.01}$ &5.53$_{-0.01}^{+0.01}$ &5.57$_{-0.01}^{+0.01}$ \\
        
        & $F_{gauss}$ ($\times 10^{-9}$) &   0.68$_{-0.1}^{+0.11}$ &0.86$_{-0.11}^{+0.11}$ &1.03$_{-0.09}^{+0.09}$ &1.01$_{-0.09}^{+0.09}$ &0.52$_{-0.06}^{+0.06}$ &0.68$_{-0.08}^{+0.08}$ &0.92$_{-0.11}^{+0.11}$ &2.53$_{-0.11}^{+0.11}$ &1.32$_{-0.18}^{+0.18}$ &2.49$_{-0.13}^{+0.13}$ \\

        & $L_{\text{total}}$ ($\times 10^{38}$) &  2.36$_{-0.49}^{+0.43}$ &2.39$_{-0.5}^{+0.43}$ &2.35$_{-0.49}^{+0.42}$ &2.18$_{-0.45}^{+0.39}$ &2.04$_{-0.42}^{+0.37}$ &2.36$_{-0.49}^{+0.42}$ &8199.01$_{-4429.05}^{+1478.49}$ &3.01$_{-0.67}^{+0.6}$ &3.48$_{-0.72}^{+0.63}$ &3.84$_{-0.87}^{+0.78}$ \\
        \hline 
    \end{tabular}
    \\
        $^*$ represents frozen parameters.\\
        $R_{bb}$ is calculated assuming a source distance of 2.13$^{+0.26}_{-0.21}$ kpc. \\
        Flux values in $\mathrm{erg\,cm^{-2}\,s^{-1}}$ are calculated in the 0.5-15.0 keV energy range using \texttt{cflux} command in \texttt{XSPEC}.\\
        Total luminosity is given in ergs s$^{-1}$ unit.\\
\end{table}
\end{landscape}

\begin{landscape}
\begin{table}
\renewcommand{\arraystretch}{1.4}
    \caption{The best-fit parameters from spectral fitting using Model-5. The free parameter for the \textit{diskbb} component are the inner disk temperature ($kT_{in}$), and the model normalization ($N_{disk}$). The energy index ($\alpha$), the electron temperature ($kT_e$), the blackbody seed photon temperature ($kT_{s}$) and the model normalization ($N_{Comptb}$) are the free parameters for the \textit{Comptb} component. The line energy ($E$), width ($\sigma$) and the normalizations ($N_{gauss}$) for the two separate Gaussian components are reported with subscript 1 and 2 for Gaussian-1 and 2 respectively. All quoted uncertainties correspond to 1$\sigma$ (68\% confidence) errors.}
    \label{table_best_fit_model5}
    \begin{tabular}{llccccccccccc}
        \hline
        \multicolumn{9}{c}{Model-5: \texttt{const*tbabs*(diskbb+Comptb+Gauss+Gauss)}}\\
        \hline
        Model & Parameter & HB1 & HB2 & NB1 & NB2 & NB3 & FB1 & FB2 & FB3 & FB4 & FB5\\
        \hline
        \texttt{Const} & 2$\times$2 & 1 & 1 & 1 & 1 & 1 & 1 & 1 & 1 & 1 & 1 \\  

        & 3$\times$3 & 1.091$\pm$0.001 & 1.091$\pm$0.001 & 1.090$\pm$0.002 & 1.095$\pm$0.001 &  1.099$\pm$0.001 & 1.093$\pm$0.001 & 1.108$\pm$0.001 & 1.090$\pm$0.01 & 1.091$\pm$0.01& 1.092$\pm$0.01\\ 

        \texttt{tbabs} & $N_{H} (\times 10^{22} \text{cm}^{-2}) $  & 0.14$^{*}$ &0.14$^{*}$ &0.15$_{-0.01}^{+0.01}$ &0.15$_{-0.01}^{+0.01}$ &0.13$_{-0.01}^{+0.01}$ &0.14$_{-0.01}^{+0.01}$ &0.14$^{*}$ &0.13$_{-0.01}^{+0.01}$ &0.13$_{-0.01}^{+0.01}$ &0.14$_{-0.01}^{+0.01}$ \\
        
        \texttt{diskbb} & $kT_{\rm in}$ (keV) & 0.85$_{-0.04}^{+0.05}$ &0.82$_{-0.06}^{+0.05}$ &0.78$_{-0.04}^{+0.05}$ &0.70$_{-0.03}^{+0.03}$ &0.61$_{-0.03}^{+0.03}$ &0.63$_{-0.02}^{+0.02}$ &0.68$_{-0.02}^{+0.02}$ &0.68$_{-0.02}^{+0.04}$ &0.72$_{-0.06}^{+0.06}$ &0.8$_{-0.08}^{+0.08}$ \\
        
        & $N_{\rm disk}$ ($\times 10^4$) & 1.6$_{-0.3}^{+0.2}$ &1.7$_{-0.4}^{+0.3}$ &2.1$_{-0.3}^{+0.5}$ &2.9$_{-0.5}^{+0.4}$ &4.5$_{-0.7}^{+0.7}$ &4.5$_{-0.4}^{+0.4}$ &3.4$_{-0.4}^{+0.3}$ &3.4$_{-0.5}^{+0.6}$ &3.0$_{-0.8}^{+0.6}$ &2.2$_{-0.8}^{+0.5}$ \\
        
        & $R_{in}$ (km) & 32.0$_{-5.0}^{+4.0}$ &34.0$_{-5.0}^{+4.0}$ &37.0$_{-5.0}^{+5.0}$ &43.0$_{-6.0}^{+5.0}$ &54.0$_{-8.0}^{+7.0}$ &54.0$_{-7.0}^{+6.0}$ &47.0$_{-6.0}^{+5.0}$ &47.0$_{-7.0}^{+6.0}$ &44.0$_{-8.0}^{+6.0}$ &38.0$_{-8.0}^{+6.0}$ \\
        
        \texttt{Comptb} & $\alpha$ & 1.42$_{-0.37}^{+0.27}$ &1.35$_{-0.27}^{+0.14}$ &1.49$_{-0.18}^{+0.14}$ &1.55$_{-0.14}^{+0.11}$ &1.61$_{-0.11}^{+0.09}$ &1.36$_{-0.12}^{+0.1}$ &1.25$_{-0.17}^{+0.13}$ &1.03$_{-0.17}^{+0.11}$ &0.79$_{-0.17}^{+0.09}$ &0.74$_{-0.14}^{+0.24}$ \\
        
        & $kT_e$ (keV) & 4.44$_{-7.38}^{+0.79}$ &3.58$_{-0.72}^{+0.28}$ &3.79$_{-0.54}^{+0.32}$ &3.59$_{-0.4}^{+0.25}$ &4.36$_{-0.27}^{+0.18}$ &2.66$_{-0.13}^{+0.1}$ &2.54$_{-0.14}^{+0.1}$ &2.44$_{-0.09}^{+0.07}$ &2.33$_{-0.09}^{+0.05}$ &2.52$_{-0.52}^{+0.12}$ \\
        
        & $kT_{\rm s}$ (keV) & 1.3$_{-0.14}^{+0.16}$ &1.15$_{-0.13}^{+0.11}$ &1.1$_{-0.08}^{+0.08}$ &1.03$_{-0.06}^{+0.05}$ &0.96$_{-0.04}^{+0.03}$ &1.01$_{-0.04}^{+0.04}$ &1.1$_{-0.06}^{+0.06}$ &1.07$_{-0.09}^{+0.08}$ &1.06$_{-0.15}^{+0.13}$ &1.31$_{-0.26}^{+0.23}$ \\
        
        & $N_{\rm Comptb}$ & 2.81$_{-0.16}^{+0.11}$ &2.84$_{-0.03}^{+0.04}$ &2.86$_{-0.02}^{+0.02}$ &2.71$_{-0.02}^{+0.02}$ &2.62$_{-0.01}^{+0.01}$ &3.02$_{-0.04}^{+0.04}$ &3.38$_{-0.08}^{+0.08}$ &3.87$_{-0.14}^{+0.12}$ &4.26$_{-0.28}^{+0.25}$ &5.24$_{-0.64}^{+0.52}$ \\
        
        \texttt{Gaussian-1} & $E_{1}$ (keV) & 6.68$_{-0.02}^{+0.02}$ &6.69$_{-0.01}^{+0.01}$ &6.71$_{-0.01}^{+0.01}$ &6.7$_{-0.01}^{+0.01}$ &6.7$_{-0.02}^{+0.02}$ &6.7$_{-0.01}^{+0.02}$ &6.69$_{-0.01}^{+0.01}$ &6.71$_{-0.01}^{+0.01}$ &6.7$_{-0.01}^{+0.02}$ &6.69$_{-0.02}^{+0.02}$ \\
        
        & $\sigma_1$ (keV) & 0.18$_{-0.02}^{+0.02}$ &0.2$_{-0.02}^{+0.02}$ &0.21$_{-0.02}^{+0.01}$ &0.21$_{-0.01}^{+0.01}$ &0.23$_{-0.02}^{+0.02}$ &0.23$_{-0.01}^{+0.02}$ &0.19$_{-0.02}^{+0.02}$ &0.19$_{-0.02}^{+0.02}$ &0.2$_{-0.02}^{+0.02}$ &0.19$_{-0.02}^{+0.02}$ \\
        
        & $N_{\rm gauss}$ & 0.13$_{-0.01}^{+0.01}$ &0.14$_{-0.01}^{+0.01}$ &0.15$_{-0.01}^{+0.01}$ &0.15$_{-0.01}^{+0.01}$ &0.19$_{-0.01}^{+0.02}$ &0.23$_{-0.01}^{+0.02}$ &0.21$_{-0.02}^{+0.02}$ &0.22$_{-0.02}^{+0.02}$ &0.24$_{-0.02}^{+0.02}$ &0.23$_{-0.02}^{+0.02}$ \\

        \texttt{Gaussian-2} & $E_{2}$ (keV) & 7.65$_{-0.07}^{+0.08}$ &7.72$_{-0.09}^{+0.1}$ &7.69$_{-0.05}^{+0.06}$ &7.66$_{-0.06}^{+0.07}$ &7.63$_{-0.08}^{+0.2}$ &7.68$_{-0.06}^{+0.11}$ &7.6$_{-0.07}^{+0.1}$ &7.51$_{-0.14}^{+0.15}$ &7.62$_{-0.08}^{+0.09}$ &7.64$_{-0.07}^{+0.09}$ \\
        
        & $\sigma_2$ (keV) & 0.3$_{*}$ &0.31$_{-0.12}^{+0.08}$ &0.32$_{-0.11}^{+0.07}$ &0.31$_{-0.13}^{+0.07}$ &0.40$_{-0.23}^{+0.10}$ &0.33$_{-0.17}^{+0.07}$ &0.40$_{-0.11}^{+0.08}$ &0.41$_{-0.12}^{+0.13}$ &0.3$^{*}$ &0.30$_{-0.11}^{+0.07}$ \\
        
        & $N_{\rm gauss}$ ($\times 10^{-2}$) & 0.48$_{-0.1}^{+0.1}$ &0.35$_{-0.08}^{+0.08}$ &0.48$_{-0.07}^{+0.07}$ &0.46$_{-0.06}^{+0.06}$ &0.63$_{-0.38}^{+0.14}$ &0.64$_{-0.25}^{+0.12}$ &0.94$_{-0.25}^{+0.18}$ &1.08$_{-0.33}^{+0.29}$ &0.78$_{-0.14}^{+0.14}$ &0.76$_{-0.25}^{+0.21}$ \\

        & F-test$_p$ & $6.5 \times 10^{-6}$  & $8.9 \times 10^{-6}$  & $4.3 \times 10^{-12}$  & $1.2 \times 10^{-11}$  &  $9.4 \times 10^{-12}$  & $1.8 \times 10^{-12}$   &  $2.1 \times 10^{-15}$  &  $9.9 \times 10^{-9}$  &  $3.7 \times 10^{-8}$  &  $4.0 \times 10^{-4}$  \\

        $\chi^2/dof$  &  & 340/362 & 427/376 & 382/386 & 346/378 & 309/375 & 349/369 & 309/369 & 336/366 & 319/371 & 343/357 \\

        \hline
        
        & $\tau$   & 6.05$\pm$12.29 & 7.87$\pm$2.34 & 6.41$\pm$1.37 & 6.1$\pm$1.07 & 4.47$\pm$0.81 &  8.46$\pm$0.93 & 9.23$\pm$1.47 & 10.89$\pm$1.9 & 13.42$\pm$3.0 & 12.42$\pm$3.03 \\

        & $y$    & 1.27$\pm$5.16 & 1.44$\pm$0.94 & 1.22$\pm$0.52 & 1.15$\pm$0.39 & 1.1$\pm$0.28 &  1.49$\pm$0.33 & 1.69$\pm$0.54 & 2.27$\pm$0.79 & 3.29$\pm$1.47 & 2.58$\pm$4.97 \\
        
        & $R_{W}$ (km)   & 17.0$_{-18.0}^{+18.0}$ &20.0$_{-10.0}^{+9.0}$ &23.0$_{-9.0}^{+9.0}$ &26.0$_{-9.0}^{+8.0}$ &30.0$_{-9.0}^{+8.0}$ &26.0$_{-7.0}^{+7.0}$ &22.0$_{-7.0}^{+7.0}$ &23.0$_{-8.0}^{+8.0}$ &22.0$_{-10.0}^{+9.0}$ &17.0$_{-14.0}^{+13.0}$ \\

        & $F_{disk}$ ($\times 10^{-7}$) &   1.32$_{-0.003}^{+0.003}$ &1.286$_{-0.002}^{+0.002}$ &1.186$_{-0.002}^{+0.002}$ &1.043$_{-0.002}^{+0.002}$ &0.849$_{-0.002}^{+0.002}$ &0.982$_{-0.002}^{+0.002}$ &1.097$_{-0.002}^{+0.002}$ &1.101$_{-0.003}^{+0.003}$ &1.211$_{-0.003}^{+0.003}$ &1.404$_{-0.003}^{+0.003}$ \\
        
        & $F_{Comp}$ ($\times 10^{-7}$) &   3.229$_{-0.004}^{+0.004}$ &3.097$_{-0.003}^{+0.003}$ &3.113$_{-0.003}^{+0.003}$ &2.969$_{-0.003}^{+0.003}$ &2.879$_{-0.003}^{+0.003}$ &3.333$_{-0.003}^{+0.003}$ &3.679$_{-0.004}^{+0.004}$ &4.414$_{-0.005}^{+0.005}$ &5.151$_{-0.005}^{+0.005}$ &5.636$_{-0.006}^{+0.006}$ \\

        & $F_{gauss}$ ($\times 10^{-9}$)   & 0.59$_{-0.1}^{+0.1}$ &0.43$_{-0.08}^{+0.08}$ &0.59$_{-0.07}^{+0.07}$ &0.56$_{-0.06}^{+0.06}$ &0.77$_{-0.07}^{+0.07}$ &0.78$_{-0.08}^{+0.08}$ &1.15$_{-0.11}^{+0.11}$ &1.31$_{-0.11}^{+0.11}$ &0.96$_{-0.14}^{+0.14}$ &0.93$_{-0.13}^{+0.13}$ \\

        & $L_{\text{total}}$ ($\times 10^{38}$)  & 2.48$_{-0.52}^{+0.45}$ &2.39$_{-0.5}^{+0.43}$ &2.35$_{-0.49}^{+0.42}$ &2.19$_{-0.45}^{+0.39}$ &2.04$_{-0.42}^{+0.37}$ &2.36$_{-0.49}^{+0.42}$ &2.61$_{-0.54}^{+0.47}$ &3.01$_{-0.65}^{+0.57}$ &3.47$_{-0.72}^{+0.63}$ &3.84$_{-0.85}^{+0.76}$ \\
        \hline
        \end{tabular}
        \\
        $^*$ represents frozen parameters.\\
        $R_{in}$ is calculated assuming a source inclination of 46$^{\circ}$ and a distance of 2.13$^{+0.26}_{-0.21}$ kpc. \\
        Flux values in $\mathrm{erg\,cm^{-2}\,s^{-1}}$ are calculated in the 0.5-15.0 keV energy range using \texttt{cflux} command in \texttt{XSPEC}.\\
        Total luminosity is given in ergs s$^{-1}$ unit.\\
\end{table}
\end{landscape}

\subsubsection{Model-5}

The spectral fit in three segments of the Z-track (HB1, NB1 and FB1) using the Model-5 along with the residual are displayed Figure \ref{spec_fit_model5}.
The different spectral components are also marked in the Figure. 
The best-fit spectral parameters for the model are reported in Table \ref{table_best_fit_model5} and the evolution of these parameters along the Z-track is illustrated in right panel of Figure \ref{spec_params_evol}.
The fit resulted in similar best-fit parameter values and trend in the variation of both disk and Comptonization parameters on comparison with the spectral fitting using Model-1.
A decrease in $kT_{in}$ followed by an increase in $R_{in}$ is observed in the HB and NB towards SA. This decrease in $kT_{in}$ is also correlated with the disk flux in NB and HB.
The seed photon temperature ($kT_s$) shows a decreasing trend correlated with $kT_{in}$ in NB and HB towards the soft apex.
The energy index ($\alpha$) changes from $\sim$ 1.35 in the beginning of HB to $\sim$ 1.65 at the NB3.
The optical depth ($\tau$) is decreases as the source approach the NB3 segment, reaches its minimum value of $\sim$5 at NB3 and then rises along the FB from $\sim$ 8 in FB1 to $\sim$12.5 in FB5.

Figure \ref{disc_params_evol} shows the evolution of the inner disk temperature ($kT_{in}$), radius ($R_{in}$) and the disk luminosity as a function of total luminosity ($L$). 
We consider a distance of 2.13$^{+0.26}_{-0.21}$ kpc for the source (\citealt{2021MNRAS.502.5455A}) to estimate the total luminosity.
The total luminosity is calculated as the sum of contribution to luminosity by the disc, the Comptonization and the Gaussian components in the spectra.
The $kT_{in}$ decreases from a higher value of $\sim$ 0.85 keV at the beginning of NB towards the soft apex and eventually increases to a higher value of $\sim$ 0.8 keV at the end of FB, where $R_{in}$ is minimum.
This is also associated with an increase in the disk flux along the FB away from the soft apex.

Again the source spectra are dominated by the Comptonization component.
The evolution of the Comptonization parameters as a function of the total luminosity for Model-5 is shown in Figure \ref{nthcomp_spec_evol}.
The Comptonization flux shows a decrease in the NB towards the SA which is correlated with the trend shown by disk flux.
The variation shown by the spectral components of both the \textit{diskbb} and \textit{comptb} components reveres as the source enters the FB from NB.
In FB, a steep increase in the Comptonization flux and disk flux is seen.
This rise in the hard counts during flaring is in contrast to what has been observed in the dipping Z-source GX 340+0, where a a slight increase or decline in hard count is seen during flaring \citet{1989A&A...225...79H}.
$\alpha$ shows significant decrease from SA to the end of FB suggesting a spectral hardening in FB. 
This nature of the source was studied previously by \citet{2014ApJ...789...98T} using \textit{RXTE} data.

To test the significance of the variability of key parameters along the Z-track for Model-5, we performed a statistical $\chi^2$ test.
We fitted the parameter evolution with a constant model, taking the weighted mean as the best-fit value. 
For parameters with asymmetric uncertainties, we adopted the average of the upper and lower errors and computed the reduced $\chi^2$ to evaluate the null hypothesis that the parameter remains constant across all segments.
A weight is given to each parameter value based on the error bar.
The result of this test is reported in Table \ref{Param_var_chi2}

\begin{table}
    \centering
    \renewcommand{\arraystretch}{1.4}
    \begin{tabular}{ccc}
    \hline
    \hline
       Parameter   & $\chi2/dof$ & p-value \\
    \hline
       $kT_{in}$    &  4.94 & $1.1\times 10^{-6}$  \\
       $R_{in}$     &  1.89 &  $4.7\times 10^{-2}$ \\
       $\alpha$     &  4.91 & $1.3\times 10^{-6}$  \\
       $kT_e$       &  7.34 & $8.7\times 10^{-11}$  \\
       $kT_{bb}$    &  1.36 & $1.9\times 10^{-1}$  \\
       $\tau$       &  3.05 & $1.2\times 10^{-3}$  \\
    \hline
    \hline
    \end{tabular}
    \caption{Results ($\chi^2/dof$) for 10 degrees of freedom of constant-model fits to the spectral parameters measured across different branches of the Z-track.}
    \label{Param_var_chi2}
\end{table}

The emission line properties also shows changes along different branches.
We performed an \texttt{F-test} in \texttt{XSPEC} to assess the significance of including the $K_\beta$ emission line in the spectral model. The test yielded a significance level more than 99\% in all segments, indicating a significant improvement to the spectral fit. The F-test results are included in Table \ref{spec_fit_model5}.
The line energies of $K_\alpha$ and $K_\beta$ appears to be unchanged and stays within error bars at $\sim$ 6.7 keV and $\sim$ 7.6 keV during the complete movement of the Z track.
However, the line width as well as the normalization shows variations. 
The line normalization of both the Gaussian lines increases monotonically from the beginning of the HB (0.13 keV for $K_\alpha$) to the end of FB (0.24) reaching an almost double value at the end of FB compared to the beginning of HB. 
However, the width ($\sigma$) increases towards soft apex and then decreases in FB. 
The maximum width of the line (0.23 keV) is observed in the NB3 segment where the disk is at a maximum distance (54 km).

\subsection{Temporal properties}

\begin{figure*}
    \centering
    \includegraphics[width=0.32\linewidth]{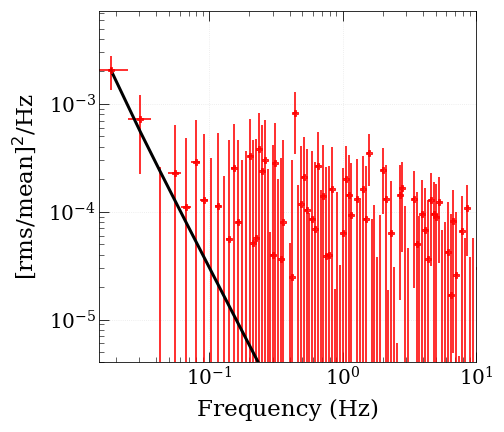}
    \includegraphics[width=0.32\linewidth]{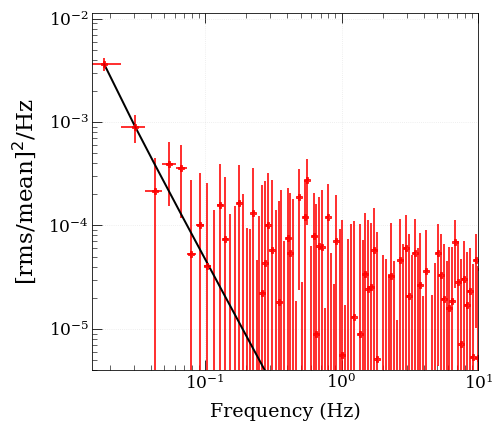}
    \includegraphics[width=0.32\linewidth]{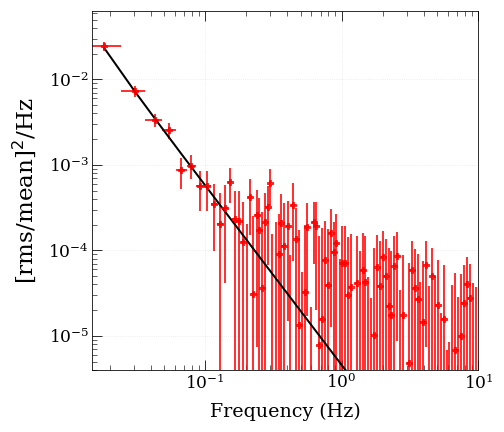}
    \caption{Power-law fit to the PDS in the HB, NB and FB branches of the Z track.}
    \label{pds_fit_plots}
\end{figure*}

The PDS generated from all branches of the Z-track were modeled using a power-law function of the form $A\nu^{-\alpha}$.
Due to the negligible signal strength above 10 Hz, the PDS fitting was restricted to the frequency range up to 10 Hz. 
The resulting fits are shown in Figure \ref{pds_fit_plots}, and the best-fit parameter values are listed in Table \ref{pds_table}. 
The root-mean-square (rms) amplitude was calculated over the broader frequency range of 10 mHz to 50 Hz. 
Among the branches, FB exhibits a higher rms amplitude compared to the other branches. 
It is observed that the RMS amplitude increases along the Z-track, with upper limits of 0.8\% in the HB and 0.96\% in the NB, reaching up to 2.5\% in the FB.
No quasi-periodic oscillations (QPOs) were detected in any of the branches.

\begin{table}
    \renewcommand{\arraystretch}{1.4}
    \centering
    \begin{tabular}{lccc}
    \hline
    \hline
    Parameter & HB   &   NB   &   FB  \\
    \hline
    $ \alpha$  & 2.4$^{+0.9}_{-1.9}$ &   2.5$^{+0.3}_{-0.4}$ &   2.1$^{+0.1}_{-0.1}$  \\
    $A$    & $<$ 0.43 &  $<$ 0.65 &   4$^{+1}_{-2}$  \\
    $rms$ (\%) & $<$ 0.96 & $<$ 0.80 & 2.6$^{+0.8}_{-0.7}$ \\
    \hline
    \end{tabular}
    \caption{The best-fit parameters of the power-law fit to the PDS in HB, NB and FB branches of the Z-track. The upper-limits and error values are reported in 68\% confidence level.}
    \label{pds_table}
\end{table}

\section{Discussion}\label{sum}

\subsection{Spectral behavior of the source}

In this work, we discuss the results from detailed spectral and timing studies of the bright NS-LMXB Sco X-1 using \textit{XSPECT} observations conducted in August 2024. 
The source light curve reveals flaring events during the observations and the CCD indicate that the observations cover the complete Z-track including HB, NB and FB.
Spectra extracted for ten different parts of the CCD are used for understanding the evolution of the source along the Z-track.
Multiple approaches are adopted to model the X-ray spectra in all ten segments.
The spectral fitting indicates that the soft thermal component in all spectra along the Z-track can be adequately modeled using either a multicolor disk blackbody (\textit{diskbb}) or a simple blackbody (\textit{bbodyrad}) component.
The hard component in the spectra could be described using a thermal Comptonization component (\textit{nthComp} or \textit{Comptb}).
Although these three models (Model-1, 2 and 5) provided a good description of the source spectra, the recent polarimetric studies using IXPE observations of Sco X-1 favor a scenario in which the soft X-ray emission arises from a multicolor disk, while the hard component originates from the Comptonization of blackbody seed photons emitted by the NS BL/SL (\citealt{2024ApJ...960L..11L}).
This idea is consistent with the Model-1 and Model-5 in our study. 
Hence we adopt Model-1 and Model-5, which used \textit{diskbb} as the best fit model.
Since, Model-5 uses \textit{Comptb} model which is more sophisticated, we fix our best fit model to be Model-5.
Apart from the continuum emission, we also observe $K_\alpha$ and $K_\beta$ emission lines of the Fe XXV ion in the spectra.
The study suggest a strong correlation of spectral properties of the source with the position in CCD.
The evolution of spectral and temporal parameters along the Z-track provides a potential explanation for the source movement along the track.
Although such studies have been conducted in the past \citep{2021MNRAS.500..772D, 2003A&A...405..237B, 2012A&A...546A..35C}, a comprehensive understanding of the full Z-track evolution in the soft X-ray band remains undisclosed.
In this work, the evolution of the nature of accretion disk and the NS blackbody emission modified by the hot electron corona are investigated in detail.
The results suggest that the emission from the accretion disk is observed directly and provide valuable information on the dynamics of accretion disk along the Z-track.

\subsubsection{Change in luminosity along the Z-track}

A conventionally accepted understanding is that the mass accretion rate ($\dot{M}$) increases monotonically along the Z-track in the direction from the the HB to NB and then to FB \citep{1989A&A...225...79H}. This view is supported by a multi-wavelength observational campaign of Cyg X-2, which shows that the UV flux increases consistently as the X-ray emission progresses along the Z-track in HB $\rightarrow$ NB $\rightarrow$ FB direction. 
In our analysis, we see that the observed total luminosity of the source in all parts of the Z-track are largely contributed by Comptonized emission, irrespective of the spectral models. 
In such a scenario, we expect the luminosity to increase along the Z-track in this direction as an increase in $\dot{M}$ would cause an increase in the soft photon supply.
However, we observe a significant decrease of $\sim$ 30\% in the total luminosity along the HB and NB in the direction HB $\rightarrow$ SA. 
This behavior has been reported previously for Sco X-1 \citep{2003A&A...405..237B, 2012A&A...546A..35C}.
A decrease in disk flux is seen in HB and NB towards the soft apex before it starts increasing along the FB.
This is correlated with the behavior of Comptonized flux in all the branches.
The Luminosity of the source in the NB is slightly higher than the Eddington limit for a 1.4 $M_{\odot}$ neutron star for which $L_{Edd} \sim 2\times 10^{38}$ erg/s and it reaches a super Eddington value of $\sim 4.0 \times 10^{38}$ ergs/s at the upper part of the FB.
This behavior of the source is consistent with the previous studies (\citealt{2003A&A...405..237B}).

\subsubsection{Behavior of Iron line emissions.}

Emission lines corresponding to $K_\alpha$ and $K_\beta$ transitions of Fe XXV ion are detected in all segments of the Z-track. 
These emission features are previously reported in the source by \citet{2021A&A...654A.102M}.
The properties of both the lines (width ($\sigma$) and normalization) are found to be correlated with the source position in the CCD.
This behavior is expected, as the mass accretion rate and the nature of disk–corona interactions are known to vary along different branches of the CCD \citep{1989A&A...225...79H, 2018MNRAS.477.5437A}.
The results may also suggest that variations in the inner disk radius possibly driven by changes in the mass accretion rate or boundary layer dynamics play a significant role in shaping the observed line profiles.

Both the $K_\alpha$ and $K_\beta$ flux shows a monotonic increase toward the FB, reaching nearly double its value at the end of FB compared to the beginning of the HB. It is evident that both emission lines consistently exhibit higher flux in FB than in HB and the NB, as shown in Figure \ref{gauss_spec_evol}. During the flaring branch, this increase is correlated with the rise in Comptonization flux and disk flux may indicate strong illumination of the inner accretion disk by the Comptonized component, leading to higher line flux. 

\subsubsection{Z-track evolution in Sco X-1}
The spectral parameters associated with both the soft disk and hard Comptonization components exhibit significant variation along the Z-track, as shown in Figure \ref{spec_params_evol}. 
The evolution of these parameters provides insight into the physical processes driving the source's behavior along the Z-track. 
Our analysis indicates that the spectral evolution in Sco X-1 is governed by a combination of thermal and non-thermal emission processes specifically, changes in both the accretion disk and the Comptonized blackbody emission. 
The Z-track evolution in the source can be explained on the basis of the systematic evolution of the Comptonization flux, optical depth, $y$-parameter and associated changes in disk parameters along the track.
We discuss the evolution of the system across the three branches in detail.
A decrease in electron temperature and an increase in photon index is seen in the HB segment towards NB. However, both the inner disk temperature and the blackbody temperature exhibit no change.

In the normal branch, the optical depth of the corona shows a gradual decrease towards SA which is associated with an increase in the electron temperature as seen for other bright Z-sources \citep{2020Ap&SS.365...41A, 2024ApJ...977..215S}.
Assuming an increase in $\dot{M}$ in the NB $\rightarrow$ FB direction, we suspect an increase in the blackbody seed photon emission during this phase. 
This is also correlated with a increase in the inner disk radius and a decrease in the inner disk temperature.
It is seen that the blackbody seed photon radius increase in NB, which pushes the disc away which in turn result in an increase in the inner disk radius. 
In this scenario, an increased influx of soft photons from the BL/SL into the corona leads to enhanced Compton cooling.
As a result, part of the cooler coronal material may condense and settle back onto the disk. 
This mechanism has been proposed to explain the observed decrease in optical depth along the NB in NS-LMXBs \citep{2003MNRAS.346..933A, 2024ApJ...977..215S}.
This mechanism result in the reduction of Comptonized emission flux and softening of the spectra in NB towards the soft apex. 
It is seen as an increase in the photon index in NB as studied earlier in \citet{2003A&A...405..237B} for Sco X-1.
Meanwhile, the remaining coronal region becomes hotter or remains at the same temperature due to a reduction in overall density.

In the flaring brach, a decrease in the inner disk radius and an associated increase in the inner disk temperature is seen. 
This matches with the variation trend followed by blackbody seed photon radius ($R_W$) and temperature. 
This decline in $R_{in}$ during flaring is previously reported for GX 17+2 by \citet{2012ApJ...756...34L} and 
the authors showed that $\dot{M}$ is constant throughout the Z-track.
During our observations, the source exhibited an increase in the disk and Comptonization luminosity.  
This suggest that the flaring in the source is driven by both the components. 
Moreover, it is unlikely that such an increase in both the total and component fluxes could occur without an increase in $\dot{M}$.
A decrease of blackbody radius in FB has been previously studied for the Sco-like sources by \citet{2012A&A...546A..35C}.
They also consider an increase in $\dot{M}$ along FB.
During this phase of high accretion rates, and with the disk closer to the neutron star surface, the radiation pressure can push some material from the accretion disc on to the corona which can result in a higher optical depth of corona in the FB, as observed.
This increase in the optical depth enhances the strength of inverse Compton scattering, which result in a higher Comptonization flux. 
This scenario is also supported by the observed increase in the Compton $y$-parameter in the FB and the decrease in the electron temperature of the corona.
This nature is previously reported for other NS-LMXBs (eg., GX 349+1; \citealt{2003MNRAS.346..933A}, LMC X-2; \citealt{2009MNRAS.398.1352A}, GX 17+2; \citealt{2020Ap&SS.365...41A}, GX 5-1; \citealt{2024ApJ...977..215S}).
This leads to the spectral hardening in the FB.
It is also evident that a steep decline in the photon index for all the models suggest spectral hardening during flaring as studied by \citet{2014ApJ...789...98T}.
We see that these variations in the Comptonization parameters such as the photon index, optical depth and electron temperature are consistent and significant irrespective of the selection of spectral models.

The absence of QPOs in the PDS provide a strong association of QPOs with high energy emission in the source. 


\section*{Acknowledgments}
The author thank XPoSat project team, facilities team, assembly, integration, and checkout teams, and mission team for their involvement and support in enabling XSPECT payload on XPoSat mission.
Authors thank GD, SAG; DD, PDMSA and Director, URSC and Department of Physics, University of Calicut for encouragement and support to carry out this research. 
We have used software provided by the High Energy Astrophysics Science Archive Research Centre (HEASARC), which is a service of the Astrophysics Science Division at NASA/GSFC and the
High Energy Astrophysics Division of the Smithsonian Astrophysical Observatory.

\section*{Data availability}
The data used in this work will be made available through the Indian Space Science Data Centre ( \href{https://www.issdc.gov.in/}{ISSDC})



\bibliographystyle{mnras}
\bibliography{mnras_template} 




\appendix


\bsp	
\label{lastpage}
\end{document}